\documentclass{aa}  

\usepackage{graphicx}
\usepackage{txfonts}
\usepackage{amsmath}	
\usepackage{amssymb}	
\usepackage{siunitx}
\usepackage{todonotes}
\usepackage{booktabs}
\usepackage{ulem}
\usepackage{makecell}
\usepackage{pbox}
\usepackage{subcaption}
\usepackage[perpage]{footmisc}

\defcitealias{Eastman1996}{E96}
\defcitealias{Dessart2005a}{D05}
\defcitealias{Hamuy2001}{H01}
\defcitealias{Pastorello2006}{P06}
\defcitealias{Faran2014}{F14}
\defcitealias{Dessart2008}{D08}

\newcommand*\dif{\mathop{}\!\mathrm{d}}

\newcommand{\Ntrain}{780}
\newcommand{\Ntest}{225}
\newcommand{\EmuAcc}{0.64 \%}
\newcommand{\PcaReconsError}{0.26 \%}
\newcommand{\EmuNintyFivePerc}{1.2 \%}

\usepackage[draft=True]{hyperref}
\usepackage{cleveref}
\crefname{equation}{Eq.}{Eqs.}
\crefname{figure}{Fig.}{Figs.}
\crefname{section}{Sect.}{Sects.}
\crefname{table}{Table}{Tables}
\crefname{chapter}{Chapter}{Chapters}
\bibpunct{(}{)}{,}{a}{}{,} 

\begin{document}

   \title{Spectral modeling of type II supernovae}

   \subtitle{II. A machine learning approach to quantitative spectroscopic analysis}

   \author{C. Vogl\inst{1,2}
          \and
        W. E. Kerzendorf\inst{3, 4, 5}
        \and
        S. A. Sim\inst{6}
         \and
        U. M. Noebauer\inst{1, 7}
          \and
        {S. Lietzau\inst{1, 8}}
          \and
          W. Hillebrandt\inst{1}
          }

   \institute{Max-Planck-Institut f\"ur Astrophysik, Karl-Schwarzschild-Str. 1, D-85741 Garching, Germany\\
   \email{cvogl@mpa-garching.mpg.de}
   \and
   Physik Department, Technische Universit{\"a}t M{\"u}nchen, James-Franck-Str. 1, D-85741 Garching, Germany
    \and
    Center for Cosmology and Particle Physics, New York University, 726 Broadway, New York, NY 10003, USA
    \and
    Department of Physics and Astronomy, Michigan State University, East Lansing, MI 48824, USA
    \and
    Department of Computational Mathematics, Science, and Engineering, Michigan State University, East Lansing, MI 48824, USA
    \and
    Astrophysics Research Centre, School of Mathematics and Physics, Queen's University Belfast, Belfast BT7 1NN, UK  \and
    MunichRe IT 1.6.1.1, K{\"o}niginstraße 107, 80802, Munich, Germany
    \and
    Vector Informatik GmbH - Niederlassung M{\"u}nchen, Baierbrunnerstraße 23, 81379, Munich, Germany
}

   \date{}

  \abstract{
  There are now hundreds of publicly available supernova spectral time series. 
  Radiative transfer modeling of this data gives insights 
  into the physical properties of these explosions such as the composition, the
  density structure, or the intrinsic luminosity\textemdash this is invaluable 
  for understanding the supernova progenitors, the explosion mechanism, or for 
  constraining the supernova distance. 
 
  However, a detailed parameter study of the available data has been out of reach 
  due to the high dimensionality of the problem coupled with the still significant 
  computational expense.
  We tackle this issue through the use of machine-learning emulators, which are
  algorithms for high-dimensional interpolation.
  These use a pre-calculated training dataset to mimic the output of a complex 
  code but with run times orders of magnitude shorter.
  We present the application of such an emulator to synthetic type II supernova 
  spectra generated with the \textsc{tardis} radiative transfer code. The results 
  show that with a relatively small training set of {\Ntrain} spectra we can 
  generate emulated spectra with interpolation uncertainties of less than one 
  percent. We demonstrate the utility of this method by automatic spectral fitting 
  of two well-known type IIP supernovae; as an exemplary application, we determine 
  the supernova distances from the spectral fits using the 
  tailored-expanding-photosphere method. We compare our results to previous 
  studies and find good agreement. This suggests that emulation of \textsc{tardis} 
  spectra can likely be used to perform automatic and detailed analysis of many 
  transient classes putting the analysis of large data repositories within reach. 
  }


   \maketitle
%
\section{Introduction}
In recent years, improvements in instrumentation as well as the 
supply of targets have led to a tremendous increase in the volume 
of spectral data gathered for astrophysical transients of all kinds.
At the same time, public databases such as the 
\href{http://www.weizmann.ac.il/astrophysics/wiserep/}{WISeREP} archive
\citep{Yaron2012} or \href{https://sne.space/}{the Open Supernova Catalog} 
\citep{SNeSpace} have made access to this data easier 
than ever before;  \href{http://www.weizmann.ac.il/astrophysics/wiserep/}{WISeREP} 
alone provides 35484 spectra for 10809 transients.\footnote{As of the 13th 
of June 2019.}

In contrast, our tools for analyzing these large spectral datasets 
have lagged behind. We can distinguish between two sets of approaches 
for dealing with such numbers of spectra.
The first, most prevalent, approach is to 
break the spectra down to a few easily-measurable diagnostic 
properties (e.g., line absorption velocities, equivalent widths),
which are then studied for correlations \citep[e.g.,][among many]{Gutierrez2017b}.
The second approach is to use the spectra as input 
for machine learning techniques; exemplary applications include spectroscopic 
classification \citep[e.g.,][]{Yip2004} or the detection of sub-classes, for 
example, of type Ia supernovae \citep{Sasdelli2016a}.
These approaches provide information about the specific measured quantities but do 
not provide a whole picture of the transient.
Radiative transfer models have the power to infer underlying physical properties 
such as the composition and structure of the ejecta; we constrain these quantities 
by adjusting parametrized models of the emitting objects such that the simulated 
and observed spectra match.
This provides, for example, information about the progenitor systems of the 
explosion for many kinds of transients
(e.g., \citealt{Hachinger2012} for SN~Ic, \citealt{Barna2017} for SN~Iax).
The main obstacle is the high cost of radiative transfer simulations; depending on 
the complexity of the underlying code, the time needed to calculate a single
synthetic spectrum ranges between minutes and days. 
This is exacerbated by high-dimensional parameter spaces; we usually aim to 
determine a combination of various abundances, the density profile, photospheric 
temperatures and velocities \citep[e.g.,][]{Dessart2006,Baron2007}. It is
prohibitively expensive to explore this parameter space automatically with 
radiative transfer models, as needed to identify 
the parameters that best reproduce the observed spectrum;
instead, the current standard method is to optimize the agreement by 
hand, relying heavily on the expertise of the 
modeler \citep[e.g.,][]{Stehle2005,Magee2017}. 
This turns each spectroscopic analysis into an extremely time-consuming process 
that can only be done for very few objects.
For example, only for three type IIP supernova has a full spectral 
time series been modeled in non-LTE in the last 15 years: SN~1999em 
\citep{Baron2004,Dessart2006}, SN~2005cs \citep[][]{Baron2007,Dessart2008} and
SN~2006bp \citep{Dessart2008}.

One way to overcome the large computational expense of radiative transfer
models in spectral fitting is to devise a fast algorithm that mimics the code. A 
very simple implementation of such an algorithm is interpolation in a pre-computed 
Cartesian grid. High-dimensional problems, such as supernovae, however, require 
the use of more complex algorithms known as emulators. In essence, the emulator 
learns the mapping from the simulation input to the output from a set of examples; 
the simulator, in this context, is treated as a black box.
Emulators are used extensively, for example in engineering, but have, as of 
yet, not found widespread application in astrophysics. Some of the sparse use 
cases have been, the prediction of the nonlinear matter power spectrum 
\citep[][]{Heitmann2009}, stellar spectra
\citep[][]{Czekala2014}, and type Ia supernova spectra 
\citep[][]{Lietzau2017}\footnote{\href{https://doi.org/10.5281/zenodo.1312512}{https://doi.org/10.5281/zenodo.1312512}}.

In this paper, we apply emulation to perform automated quantitative spectroscopic 
analysis of type II supernovae. We 
use Gaussian-process interpolation in the PCA\footnote{See 
\cref{sec:Preprocessing} for more details on Principal Component Analysis (PCA).} 
space to reproduce the output of our radiative transfer code, a modified version 
of the Monte Carlo code \textsc{Tardis} \citep[][]{Vogl2019}. With the 
emulator, we reduce the time for the calculation of a synthetic spectrum from 
hours to milliseconds; this, in turn, makes it possible to fit spectra using 
conventional optimization methods or to explore the parameter space with a sampler.

We showcase the emulator by inferring distances to type IIP supernovae using 
the tailored-expanding-photosphere method 
\citep[tailored EPM; ][]{Dessart2006,Dessart2008}. This method uses 
spectroscopic analysis of type IIP supernovae to obtain distances with 
small uncertainties, for example, for cosmological studies.
A cosmological application requires a large number of uniformly studied 
supernovae, which will be made possible by the use of emulators.
Such an endeavor will provide an independent, physics-based probe of the cosmic 
expansion history. 

\Cref{sec:ParamModels} provides a short introduction to supernova models 
and their use for parameter inference.
\Cref{sec:SpectralLibrary} describes the library of synthetic 
spectra that forms the basis of our machine learning approach.
\Cref{sec:EmulatorMethod} is dedicated to the presentation 
of the spectral emulator: the machine learning techniques, the training 
process, and the prediction step.
In \cref{sec:EmulatorEvaluation}, we assess the predictive performance  
by comparing emulated and simulated spectra for a set of 
independent test models. We continue with the application of the emulator to the 
modeling of spectra of SN~1999em and SN~2005cs in \cref{sec:observations}. We 
show the application of measuring distances using the tailored EPM in 
\cref{sec:Distances}. \cref{sec:conclusions} summarizes the results and gives an 
outlook on the next steps.

\section{Parametrized supernova models with \textsc{Tardis}} \label{sec:ParamModels}
We use simple parametrized models of the supernova ejecta to make inferences 
about the supernova properties. In defining these models, we assume the ejecta to 
be spherically symmetric and in homologous expansion. 
This allows us to discretize the spectrum formation region into a set of shells 
that are specified by their composition, density, and expansion velocity. 
It is often useful to simplify the model specification further, for example, by 
assuming an analytic form for the density profile (e.g., a power law)
or by assuming uniform abundances\textemdash this reduces the number of 
parameters considerably. 

Since we do not simulate the creation of the radiation field 
self-consistently, we treat the radiation field at the inner boundary as a model 
parameter. Specifically, we assume a blackbody characterized by a temperature 
$T_\mathrm{inner}$; this is well motivated for type II supernovae (SNe~II)
since the continuum opacity from hydrogen leads to a full thermalization of the 
radiation field at high optical depths.

To assess if the thus defined parametrized model is consistent with the 
observations, we simulate the radiation transport through the discretized ejecta; 
then, we compare the synthetic and observed spectra. In \textsc{Tardis} 
\citep[][]{Kerzendorf2014}, we use a Monte Carlo approach based on the
indivisible energy packet scheme of
\citet{Lucy1999b,Lucy1999a,Lucy2002,Lucy2003} to this end.
The version used in this paper \citep[][]{Vogl2019} simulates the
effects of bound-bound, bound-free, free-free, as well as 
collisional interactions on the radiation field; it accounts for NLTE 
effects in the excitation and ionization of hydrogen and calculates 
the thermal structure of the envelope from the balance of heating and cooling 
processes.

\section{Creation of a SN~II spectral training set} \label{sec:SpectralLibrary}
As a first step towards the spectral emulator,
we need to calculate a set of
synthetic SNe~II spectra, which will serve as the training data.
We have selected photospheric velocity
$v_\mathrm{ph}$, photospheric temperature $T_\mathrm{ph}$, metallicity $Z$, time 
of explosion $t_\mathrm{exp}$ and
steepness of the density profile $n=-\dif{\ln{\rho}}/\dif{\ln{r}}$ as the 
parameters of our model
grid. 
The latter provides a simple parameterization of the 
density profile, which, as demonstrated, for example, by
{\citet[][]{Chevalier1976,Blinnikov2000} or \citet{Dessart2006}},
describes the outer density distribution with sufficient accuracy.

Most of the spectral evolution of photospheric phase 
SNe~II, as well as the differences between individual 
objects, can be explained by variations in the expansion 
velocity, the temperature as well as the density profile.
As such, the parameters usually considered in
quantitative spectroscopic analyses, such as those of 
\citet{Dessart2006} or \citet{Dessart2008} are $v_\mathrm{ph}$, $T_\mathrm{ph}$
and $n$.\footnote{We define the photospheric temperature $T_\mathrm{ph}$ as the 
temperature of the electron gas at an electron scattering 
optical depth of $\tau = 2/3$.} In addition to these essential 
parameters, we include 
the metallicity $Z$.\footnote{We let metallicity refer to the abundances of all 
elements except H, He, C, N, and O. 
The mass fractions of the thus defined metal species are multiples of the solar 
neighborhood values $Z_\odot$ of \citet{Asplund2009}.} 
Observed SNe~II show a wide range of metallicities 
\citep[see e.g.,][]{Anderson2016,Taddia2016} and the associated changes to the 
spectrum are significant, in particular in the blue.
For the purpose of inferring accurate distances, it is also important to 
allow for variations in the time of explosion $t_\mathrm{exp}$.
While the effects on the shape of the spectral energy distribution 
(SED) are small, $t_\mathrm{exp}$ affects 
the absolute value of the flux through the modulation of 
the photospheric density {(for any given density profile)} and therefore the 
amount of continuum flux dilution \citep[see e.g.,][]{Eastman1996}.
Other potentially relevant parameters are 
the abundances of CNO process elements, which have been 
investigated, for example, by \citet[][]{Baron2007}, as well as 
the H/He abundance ratio as studied, for example, by \citet{Dessart2006}. 
For the first demonstration of our method,
we refrain from varying these parameters and instead
adopt CNO-cycle equilibrium values for the relevant abundances from 
\citet{Prantzos1986} as in \citet{Dessart2005,Dessart2006}.\footnote{
Specifically, we adopt the following number density ratios:\\H/He = 5, N/He = 
\SI{6.8e-3}{}, C/He = \SI{1.7e-4}{}, and O/He = \SI{e-4}{}. These ratios together 
with the mass fractions of the metal species specify the composition completely.}

\Cref{tab:grid_params} lists the ranges of parameters $v_\mathrm{ph}$, 
$T_\mathrm{ph}$, $Z$, $t_\mathrm{exp}$, and $n$ covered by our model grid. 
We have chosen the parameter space such that it allows for the modeling 
of a large variety of SNe~II between roughly one and three
weeks after explosion. 

\begin{figure*}
  \begin{center}
  \hspace*{-0.45cm}
    \includegraphics[width=\textwidth]{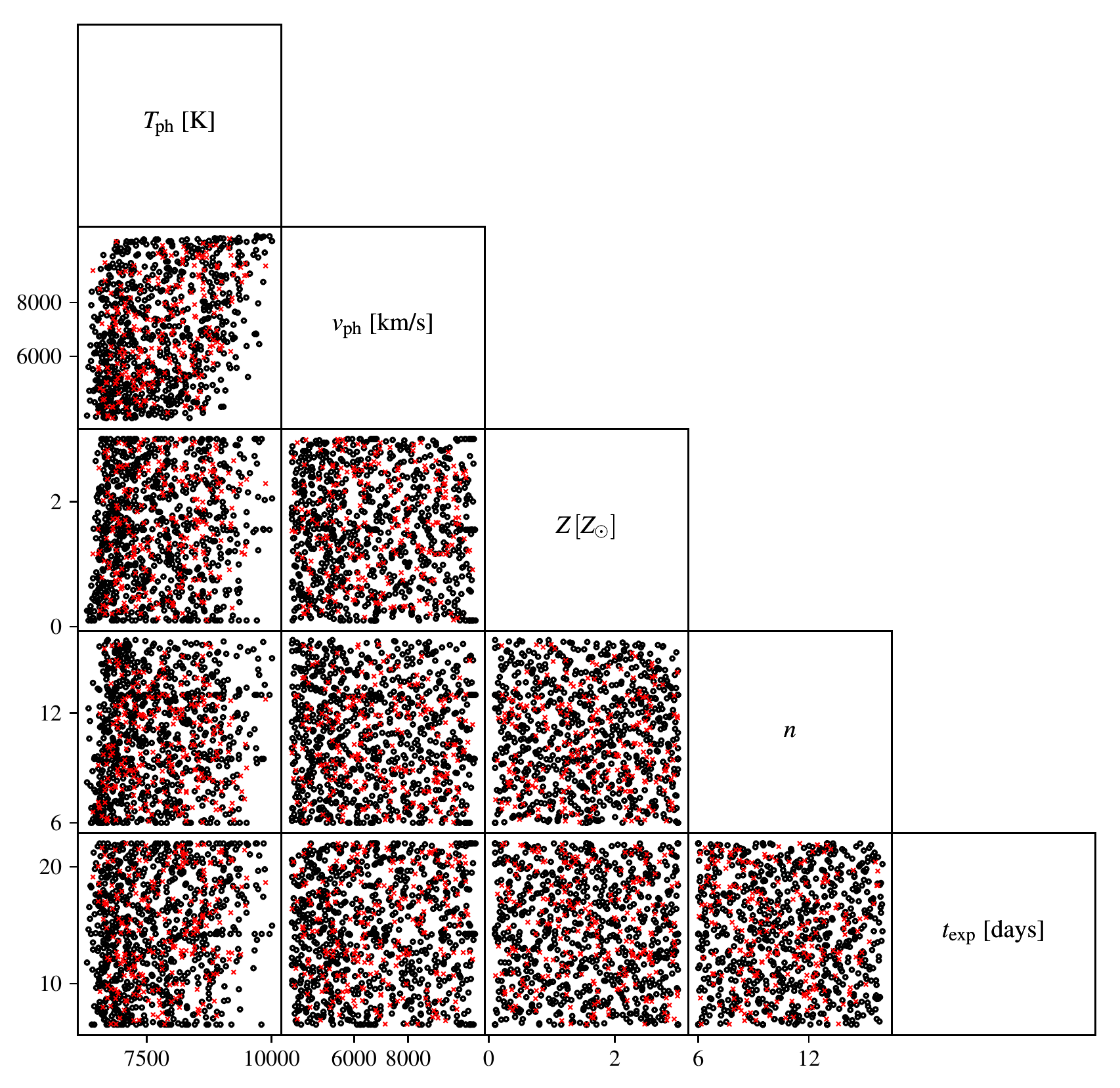}
  \end{center}
  \caption[]{Scatterplot matrix of the parameters of the training data (black)
  and test data (red) of the spectral emulator.}
  \label{fig:Grid}
\end{figure*}

In practice, we cannot directly specify $T_\mathrm{ph}$ and $v_\mathrm{ph}$
since both are emergent and not input properties of the simulation; instead, we 
use the inner boundary temperature
$T_\mathrm{inner}$, that is to say, the temperature of the injected blackbody 
radiation\footnote{Typically, the inner edge of our computational domain, where 
the packets are injected, lies at an electron scattering optical depth of around 
20.}, and a simple analytic estimate for 
the photospheric velocity $v_\mathrm{ph}^*$.
We set up a five-dimensional latin hypercube design \citep{Stein1987} in these 
parameters, which is optimized to fill the space nearly uniformly.
In the next step, we perform radiative transfer calculations
for the resulting set of {\Ntrain} models and obtain synthetic spectra 
as well as the real values of $T_\mathrm{ph}$ and $v_\mathrm{ph}$.
\Cref{fig:Grid} shows pairwise projections for the completed set of parameters 
$v_\mathrm{ph}$, $T_\mathrm{ph}$, $Z$, $t_\mathrm{exp}$, and $n$. 
The grid of models displays a slight distortion 
in the $v_\mathrm{ph}$-$T_\mathrm{ph}$ plane 
as a result of our use of $v_\mathrm{ph}^*$ and $T_\mathrm{inner}$ as 
proxies for these quantities.

A common approach in machine learning is to generate a test set in addition 
to the training set to assess the predictive accuracy. We compute {\Ntest} 
models (in addition to the {\Ntrain} training models).
The parameters for these models are sampled uniformly from the same range of
$v_\mathrm{ph}^*$, $T_\mathrm{inner}$, $Z$, $t_\mathrm{exp}$, and $n$
as covered by the training data. We include the properties of the test data
in \cref{fig:Grid} to facilitate the comparison between 
the sets of models.

One of the challenges in the setup of the model grid is deciding how large
the training set of models really needs to be.
Ideally, the number of training models should be large enough 
to guarantee that the interpolation uncertainty in the
spectra is not the dominant contribution to the error in the inferred
parameters; at the same time, the training size should be kept as small as possible
for reasons of computational expediency. 
In practice, the best possible trade-off is difficult to identify since
the conversion from the interpolation errors in the spectra to errors in the parameters is
nontrivial.
To be on the safe side, however, one can aim to have the interpolation uncertainty
significantly smaller than the systematic mismatch between model and observation; for
the parameter space we consider, this is indeed the case for the training set
we have used (see \cref{sec:EmulatorEvaluation}).

\begin{table}
  \caption[]{Parameter range covered by the spectral library.}
  \centering
  \begin{tabular}{cccccc}
   \toprule \toprule
     & $v_\mathrm{ph}$\,[km/s]  & $T_\mathrm{ph}\,$[K] & $Z\,[Z_\odot]$ & $t_\mathrm{exp}\,$[days] & n\\
    \midrule
    min & 3700 &  6300 & 0.1 & 6.5 & 6 \\
    max & 10500 &  10000 & 3.0 & 22.0 & 16 \\
    \bottomrule
  \end{tabular}
  \label{tab:grid_params}
\end{table}

\section{Spectral emulator} \label{sec:EmulatorMethod}
We use the synthetic spectra from the previous 
section to create two separate emulators: one for spectra and one for absolute
magnitudes. We set up these emulators in two steps.
First, we preprocess the training data and synthesize absolute 
magnitudes; this is followed by dimensionality reduction of the preprocessed 
spectra through PCA decomposition.
Second, we train Gaussian processes to interpolate the spectra within the 
PCA space and to predict absolute magnitudes.

\subsection{Preprocessing and dimensionality reduction} \label{sec:Preprocessing}
The synthetic spectra have a range of values that varies widely both with 
wavelength and between models. In addition, they contain non-negligible Monte 
Carlo noise.
For the successful application of machine learning techniques, it is 
crucial to preprocess the noisy, unscaled data to standardize it and to remove 
unwanted sources of variations.
We start by smoothing the spectra with a fifth-order Savitzky-Golay filter 
\citep{Savitzky1964} to reduce the effect of Monte Carlo noise.
Savitzky-Golay filtering performs well in preserving the shape of spectral 
features, even weak ones, which makes it a popular choice 
for denoising astrophysical spectra 
\citep[see e.g.,][]{Hugelmeyer2007,Poznanski2010,Sasdelli2014}.
Next, we approximately correct for the variations in the position of 
spectral features between models by Doppler shifting each spectrum by the 
photospheric velocity $v_\mathrm{ph}$. This roughly maps the  
absorption minimum of a spectral feature to the same 
wavelength for all models. Since we do not assume a distance for fitting
an observed spectrum (see \cref{sec:observations}), we can standardize the 
spectral library further
by discarding the information about the luminosity.
Specifically, we normalize the shifted synthetic spectra to have unit 
flux at \SI{6000}{\AA}; this provides a good standardization of the 
continuum flux levels between models since no strong line features form at this position.
Finally, we apply a linear transformation to the fluxes in each wavelength 
bin such that in each bin the values for the full spectral library span 
a range from -0.5 to 0.5 (see \cref{sec:MinMax}). 
For each preprocessing step, we restrict the 
considered wavelength range to the minimum range needed to model the 
observations in \cref{sec:observations}.\footnote{Note that this wavelength 
range deviates slightly from that of the observed spectra to allow for Doppler 
shifting the model spectra in the preprocessing.}
In practice, this corresponds to a wavelength window from 
roughly \SIrange{3200}{9500}{\AA}. In contrast, we only smooth the test 
spectra but do not preprocess them further: we will compare them to the 
emulated spectra in the same fashion one would do for observational data.

In the final step, we reduce the dimensionality of our spectral library. 
Each preprocessed spectrum consists of a few thousand wavelength bins, 
a number which by far exceeds
that of the physical parameters used in its creation. To obtain a less 
correlated, more compact representation of our data, we use
Principal Component Analysis (PCA).\footnote{Specifically, we use the 
probabilistic PCA model of \citet{Tipping1999} as implemented in scikit-learn 
\citep{scikit-learn}.}
PCA has been applied successfully to observed spectra of a wide range of 
astrophysical objects, including QSOs \citep[e.g.,][]{Francis1992}, stars 
\citep[e.g.,][]{Bailer1998}, galaxies \citep[e.g.,][]{Connolly1995} and 
supernovae \citep[e.g.,][]{Sasdelli2014,Williamson2019}.
The basic idea is to find an orthogonal basis for the data that is a linear 
transformation of the original but where the axes are aligned 
with the directions of maximal variance. 
Since by construction each successive principal component explains less of the 
variance in the data, we can reduce the dimensionality of our 
dataset by truncating the basis; instead of using the full set of 
$N_\mathrm{L}$ principal component eigenvectors, where $N_\mathrm{L}$ 
is the number of spectra in the spectral library, we use only the first $N < 
N_\mathrm{L}$ components. We select the dimensionality
$N$ of the truncated basis through cross-validation on the training sample 
since our main goal is the accurate prediction of synthetic spectra. The 
cross-validation performance increases, at first, as more principal components 
are included but at some point levels off when the additional components stop 
to contain meaningful information; it is at this point that we truncate 
the PCA basis. For the spectral emulator presented in this paper, this 
approach leads us to use $N=80$ principal components, which explain 99.97\% 
of the total variance; this is a significant reduction compared to the original
$\approx1500$ wavelength bins.
If necessary\footnote{For example, for sampling in high-dimensional parameter 
spaces the emulation speed may become a limitation.}, the number of principal
components can be reduced even further with only minor losses in accuracy.
By projecting each preprocessed spectrum $\hat{f}_{k}$ onto
the basis vectors $\xi_i$ of the thus truncated basis, we obtain a 
compact representation of the input data in terms of a set of $N$ principal 
component weights $\{w_{ik}\}$. From these principal component weights, we can 
lossily reconstruct every input spectrum
as a linear combination of the principal components $\xi_i$ and the mean 
spectrum $\xi_\mu = \sum_k^{N_L} \hat{f}_{k} / N_\mathrm{L}$:
\begin{equation} \label{eq:PCA}
    \hat{f}_{k} \approx \xi_\mu + \sum_{i=1}^{N} w_{ik} \xi_i
\end{equation}
For the selected number of principal components, the mean
fractional reconstruction error for this procedure is {\PcaReconsError}.

In addition to the spectral preprocessing and the dimensionality 
reduction, we synthesize Johnson-Cousins \textit{B},\textit{V}, \textit{I} 
magnitudes from the unprocessed synthetic spectra.\footnote{For our synthetic 
photometry, we use the filter functions of \citet{Bessell1990}.} These 
serve as training data for a separate emulator that predicts absolute 
photometric magnitudes for a set of model parameters 
$v_\mathrm{ph}$, $T_\mathrm{ph}$, $Z$, $t_\mathrm{exp}$, and $n$. 
This allows us, in \cref{sec:observations}, to convert the inferred 
parameters from the spectral fitting into a distance estimate based on 
the observed photometry.

\subsection{Gaussian process interpolation}
\subsubsection*{Spectra}
To predict a spectrum for a new set of input parameters \linebreak
$\theta = (v_\mathrm{ph}, T_\mathrm{ph}, Z, n, t_\mathrm{exp})$, we have 
to interpolate between the principal component weights $\{w_{ik}\}$, which 
form the compressed version of our spectral library. We choose to model the 
weights $w_i$ for each principal component $\xi_i$ independently since, by 
construction, the weights for the different components are at least 
linearly uncorrelated.
As in \citet{Czekala2014}, we use Gaussian 
processes \citep[e.g.,][]{Rasmussen06gaussianprocesses} for the 
interpolation. Gaussian processes (GPs) are a powerful, probabilistic 
tool for regression analysis, 
which are steadily gaining in popularity in the astrophysical community 
\citep[see, e.g.,][]{2015Rajpaul,2017ForemanMackey}.
As a non-parametric method, GPs offer increased flexibility for modeling 
complicated signals compared to more conventional approaches such as linear or 
polynomial regression.

Fundamentally speaking, GPs provide a generalization of the Gaussian 
probability distribution from finite-dimensional random variables 
to functions. Following this analogy, each GP is
characterized by a mean- and a covariance function.
The covariance function $k$ controls the covariance between the 
distribution of random function values at any two points $\theta, \theta^{'}$ 
in the parameter space. 
As such, it determines the properties of the 
functions that can be drawn from the GP including, for example, their 
smoothness, periodicity and so forth. In the context of regression analysis, 
the choice of the covariance function sets the prior distribution of functions 
that we expect to see in the data.\footnote{It is customary to assume a 
zero-mean for the prior distribution of possible regression functions 
\citep[see, e.g.,][]{Rasmussen06gaussianprocesses}.} 
A particularly important class of covariance functions
are the so-called stationary covariance functions, which do not 
depend on the positions $\theta, \theta^{'}$ in the input space 
but only on their distance $r = ||\theta - \theta^{'}||$.
The most commonly used members of this class include the squared exponential, 
the Mat\'{e}rn and the rational quadratic covariance function 
\citep[see, e.g.,][]{Rasmussen06gaussianprocesses,murphy2013machine}. 
The type of covariance function is a hyperparamter of the machine-learning 
approach and can, similar to the preprocessing steps, be set based on the 
cross-validation performance. After some experimentation, we have adopted 
covariance functions from the Mat\'{e}rn family:
\begin{equation} \label{eq:Matern}
    k_\mathrm{Matern}(r) = \sigma_f^{2} \frac{2^{1-\nu}}{\Gamma(\nu)} 
    \left(\sqrt{2 \nu} r\right)^{\nu} K_\nu \left(\sqrt{2 \nu} r\right)
\end{equation}
Here, $\sigma_f^{2}$ denotes the signal variance, $\nu$ is a parameter 
that regulates the smoothness of the GP, $\Gamma$ is the gamma 
function, and $K_{\nu }$ is the modified Bessel function of the second kind. 
Since we do not expect the weights, which encode our MC synthetic spectra, to 
be noise free, we include an additive contribution of homoscedastic 
white noise in the covariance function
\begin{equation}
   k(r) = k_\mathrm{Matern}(r) + \sigma_n^2 \, \delta (r).
\end{equation}
Here, $\sigma_n^2$ is the noise variance and $\delta$ the Dirac 
delta function. We complete the description of the covariance function by 
defining the distance as
\begin{equation}
    r^{2}(\theta, \theta^{'}) = (\theta - \theta^{'})^{\mathrm{T}} M \, (\theta - \theta^{'}),
\end{equation}
where $M$ can be any positive semidefinite matrix. For reasons of simplicity, 
we only consider diagonal matrices of the following type
\begin{equation}
    M =
  \begin{bmatrix}
    \frac{1}{{l_{v_\mathrm{ph}}}^2}{} & & & \\
    & \frac{1}{{l_{T_\mathrm{ph}}}^2} & &\\
    &  & \ddots &\\
    & & & \frac{1}{{l_{t_\mathrm{exp}}}^2}
  \end{bmatrix}
\end{equation}
in this paper.
For this choice of metric, each dimension of the input space 
$(v_\mathrm{ph}, T_\mathrm{ph}, Z, n, t_\mathrm{exp})$ has its own 
characteristic length-scale $(l_{v_\mathrm{ph}},\dots ,l_{t_\mathrm{exp}})$ 
for variations in the function values.

Finally, to make predictions, 
we have to move from the
prior distribution of functions 
to a posterior distribution of functions that agree with 
the training data. Mathematically speaking, this is achieved by 
conditioning the zero-mean prior GP on the observed values. 
The conditional GP has a non-zero mean function $w_i(\theta)$ that is 
determined by the values $w_{ik}$ of the training data and the covariances 
$k_i(\theta, \theta_k)$ between the location $\theta$ and the training 
locations $\theta_k$. The relevant expressions for the predictive mean and 
variance can be found in standard textbooks such as 
\citet[][their Algorithm 2.1]{Rasmussen06gaussianprocesses}.
Given a set of hyperparameters $(\sigma_n^2, \sigma_f^2, 
l_{v_\mathrm{ph}},\dots ,l_{t_\mathrm{exp}}, \nu)$, these equations yield the 
interpolated values for the principal component weight as well as an estimate 
of the interpolation uncertainty. 
The parameter $\nu$ regulating the smoothness properties of the process is 
difficult to constrain through the data; after some experimentation, we have 
adopted $\nu = 3/2$, corresponding to functions that are once mean-square 
differentiable.
We set the rest of the hyperparameters by numerically maximizing the marginal 
likelihood of the training data under the GP model. We repeat this process 
$N$ times since we model the weights $w_i$ 
for each principal component $\xi_i$
independently.

\Cref{eq:PCA} allows us to predict `preprocessed' spectra 
$\hat{f}(\theta)$ using the trained GPs.
To arrive at a spectrum that we can compare to observations, 
we have to reverse some of the preprocessing steps  used to standardize the 
input spectra for PCA in \cref{sec:Preprocessing}. This involves inverting the 
linear transformation applied to map the fluxes in each bin to the range 
$[-0.5, 0.5]$, as well as blue-shifting the spectrum by the photospheric 
velocity $v_\mathrm{ph}$.

\subsubsection*{Absolute magnitudes}
In \cref{sec:Preprocessing}, we removed the luminosity information from the 
synthetic spectra to standardize them further for PCA. We
train additional GPs for the prediction of the absolute magnitudes, which we 
need for our distance inferences in \cref{sec:observations}.

As part of the data preprocessing, we have synthesized
Johnson-Cousins \textit{S}=\{\textit{B},\textit{V},\textit{I}\} magnitudes 
$M_S$ from the unprocessed synthetic spectra. Before we use these as training 
data for the GPs, we remove the variation in the magnitudes introduced by 
differences in the physical sizes of the supernova models.
Specifically, we transform from absolute magnitudes to magnitudes at the 
position of the photosphere
\begin{equation}
  m_{S}^{\mathrm{ph}} = M_{S} +  5 \log \frac{R_\mathrm{ph}}{\SI{10}{pc}},
\end{equation}
where $R_\mathrm{ph} = v_\mathrm{ph} t_\mathrm{exp}$. We model each bandpass 
with a GP with a Mat\'{e}rn covariance function (see, \cref{eq:Matern}) that 
has a smoothness parameter $\nu=5/2$.
The hyperparameters 
$(\sigma_n^2, \sigma_f^2, l_{v_\mathrm{ph}},\dots ,l_{t_\mathrm{exp}})$ 
for the individual bandpasses are set in the same fashion as for the spectral 
emulator. Finally, to predict absolute magnitudes for a new set of input 
parameters $\theta = (v_\mathrm{ph}, T_\mathrm{ph}, Z, n, t_\mathrm{exp})$, 
we evaluate the trained GP and subtract $5 \log (R_\mathrm{ph}/ \SI{10}{pc})$.

\section{Evaluation of the emulator performance} \label{sec:EmulatorEvaluation}
To allow the reliable inference of parameters from supernova spectra, it 
is crucial that the emulator reproduces the output of our simulation 
code {\sc tardis} to high precision.
We assess the predictive performance of our method by comparing the predicted 
spectra and absolute magnitudes to a set of independently collected test data. 
Our strategy for the calculation of the {\Ntest} test models is described in 
detail in \cref{sec:SpectralLibrary} and the associated preprocessing 
procedure in \cref{sec:Preprocessing}.

\subsubsection*{Spectra}
\begin{figure*}
  \begin{center}
    \includegraphics[width=\textwidth]{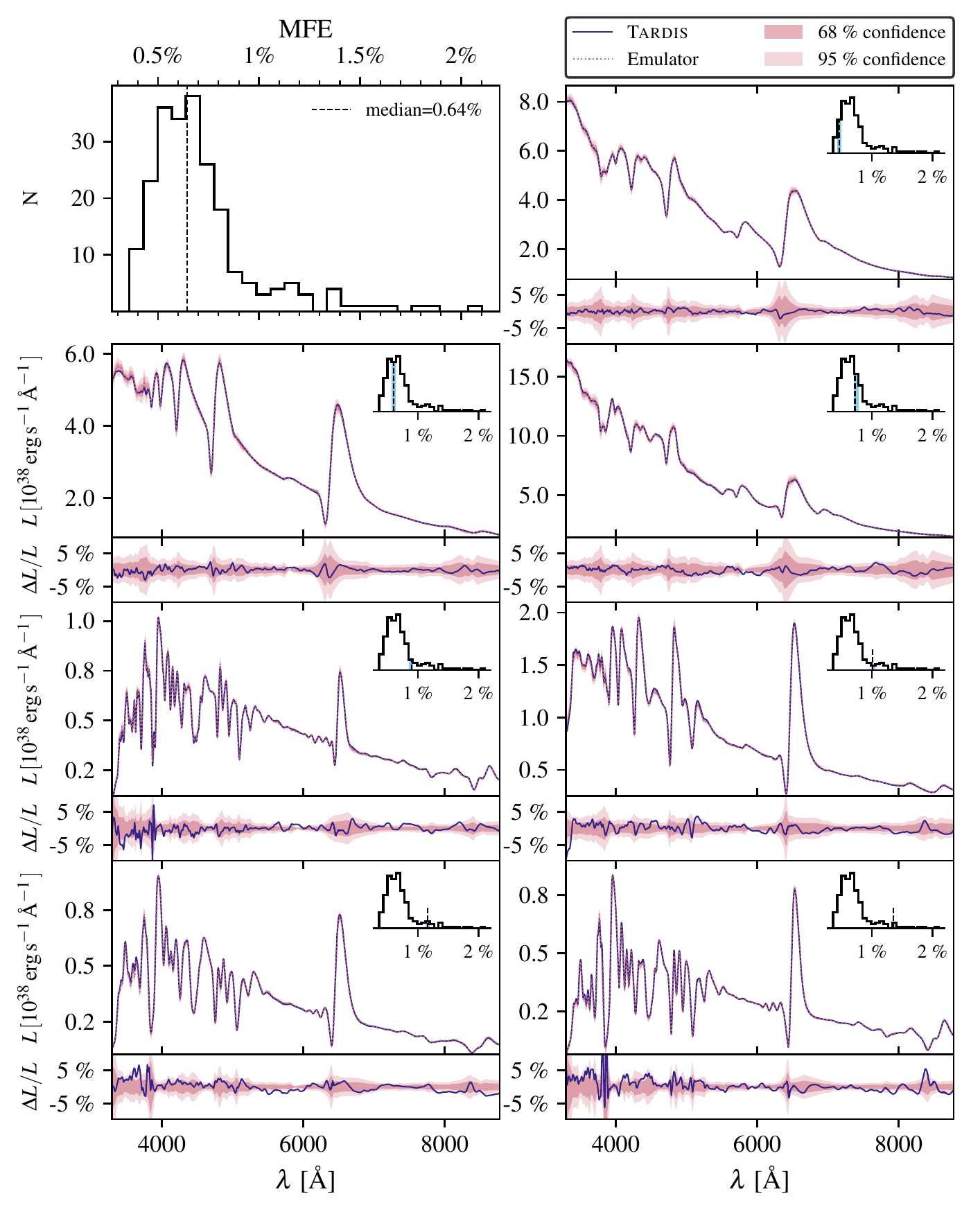}
  \end{center}
  \vspace*{-5mm}
  \caption[]{Evaluation of the emulator performance. The test performance is 
  summarized in the top left panel, which shows a histogram of the 
  test errors; specifically, the MFE (see Eq. \ref{eq:MFE}) is displayed. 
  The other panels provide a direct comparison between emulated and simulated 
  spectra for a subset of the test data. Each panel contains a histogram of 
  the test errors, in which the position of the current model 
  is highlighted. To highlight the subtle differences between the predicted 
  and true spectra, the fractional difference $\Delta L/ L$ is shown in the 
  lower section of each panel (solid blue line). In both sections, the
  shaded regions indicate the $68\%$ and $95\%$ confidence intervals for the 
  prediction of the emulator.}
  \label{fig:spectra_validation}
\end{figure*}

In \cref{fig:spectra_validation}, we compare simulated and emulated 
spectra for a subset of the test models; the selected subset approximately 
spans the range of deviations encountered in the full test data.
We have scaled the emulated spectra back to physical units for the comparison 
of spectral shapes.\footnote{As discussed in \cref{sec:Preprocessing}, we 
discard any useful luminosity information during the preprocessing of the 
training spectra; this means that it is only meaningful to compare spectral 
shapes.}
Despite covering a wide variety of spectral appearances, including, for 
example, SEDs with very broad or very narrow features, with or without line 
blanketing, the agreement is overall excellent. In addition, in most cases, 
the deviations are within the 95\% confidence interval of the emulator 
prediction with areas of larger residuals corresponding to regions with 
increased emulation uncertainties. 

\begin{figure*}
  \begin{center}
  \hspace*{-0.45cm}
    \includegraphics[width=\textwidth]{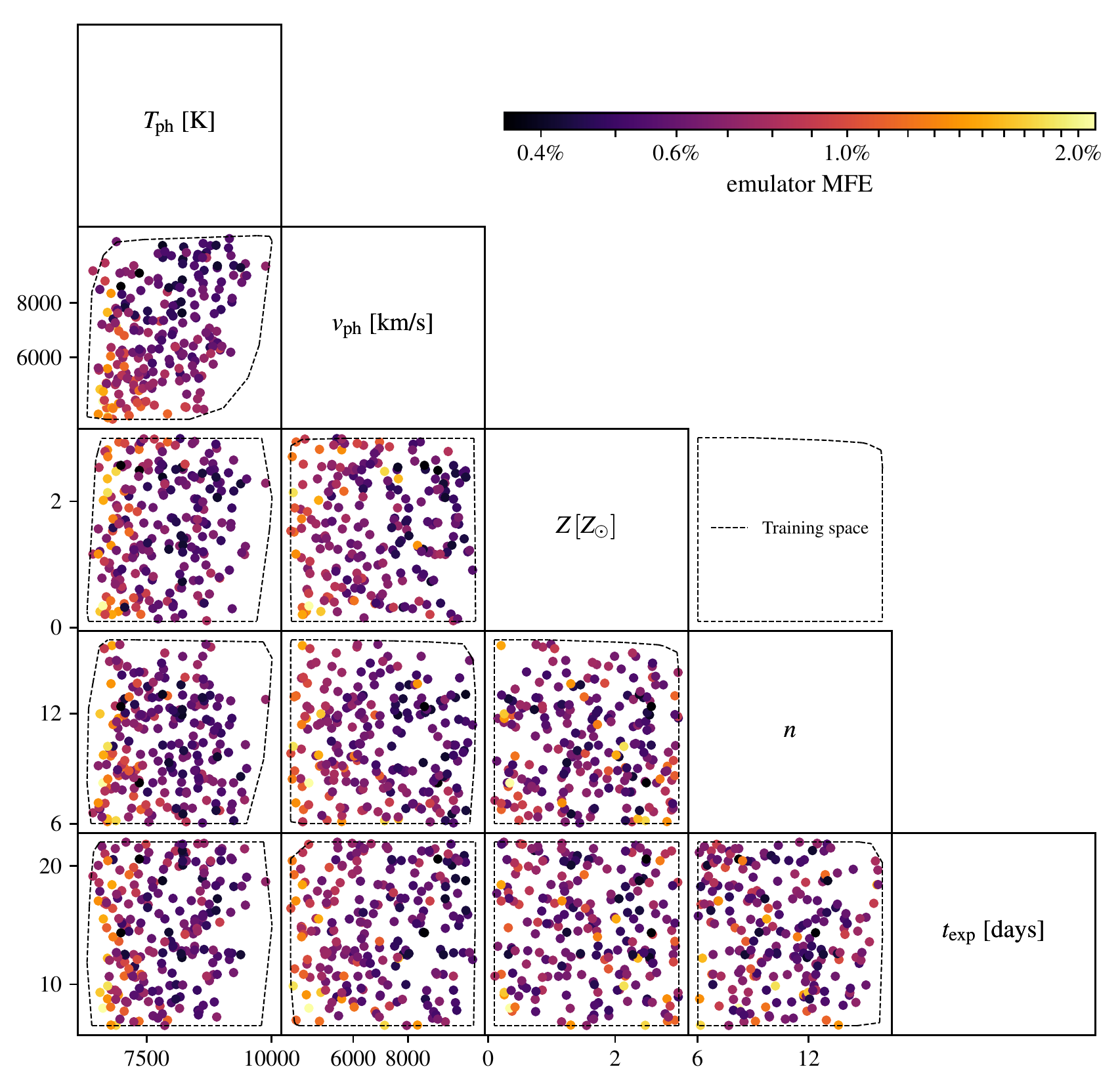}
  \end{center}
  \caption[]{Test errors for the spectral emulator as a function of the input 
  parameters. We show the color-coded MFE (see Eq. \ref{eq:MFE}) between 
  emulated and simulated spectra for all two dimensional projections of the 
  test parameters. The region enclosed by the dashed black line indicates the 
  parameter space covered by the training data.}
  \label{fig:Grid_errors}
\end{figure*}

In order to quantify the test performance, we need to define a quality 
metric that expresses the mismatch between two spectra in a single number. 
We use the mean fractional error
\begin{equation} \label{eq:MFE}
    \mathrm{MFE} = \frac{1}{N_\lambda} \sum_{i=1}^{N_\lambda} \frac{|f_{\lambda,i}^\mathrm{emu} - f_{\lambda,i}^\mathrm{test}|}{f_{\lambda,i}^\mathrm{test}},
\end{equation}
where $f_{\lambda,i}^\mathrm{test}$ and $f_{\lambda,i}^\mathrm{emu}$ are the 
test and emulated spectra respectively, and $N_\lambda$ is the number of 
wavelength bins. By using the MFE instead of, for example, the mean squared 
error, we give approximately the same weight to the red (low flux) and blue 
(high flux) parts of the spectrum. We summarize the test performance in the 
top left panel of \cref{fig:spectra_validation}, which shows a histogram of 
the MFEs for the entire test sample. The median MFE is {\EmuAcc}, confirming 
the excellent agreement found by visual inspection. For 95 percent of the test 
spectra the deviation is less than \linebreak 
{\EmuNintyFivePerc}; for the remaining 5 percent maximum differences of 
around 2 \% are possible. 
To assess how the emulator performance varies within the parameter space,
we have modified the scatterplot matrix of the test parameters to include the 
color-coded MFE (see \cref{fig:Grid_errors}). The figure demonstrates that a 
significant fraction of the cases with appreciable mismatches can be traced 
back to models 
near the edge of the training parameter space (or even outside of it). 
We also notice a slight decrease in performance towards lower velocities, 
temperatures, and higher metallicities. This trend is to be expected since the 
complexity of the SED increases in these directions of the parameter space. 
For example, in the case of velocity, we move from a few blended features to a 
forest of individual metal lines; each of these lines evolves individually, in 
a nonlinear fashion making it difficult to model the spectral evolution based 
on a PCA decomposition.

\subsubsection*{Absolute magnitudes}
\begin{figure}
  \begin{center}
  \hspace*{-0.45cm}
  \vspace{-0.45cm}
    \includegraphics[width=0.45\textwidth]{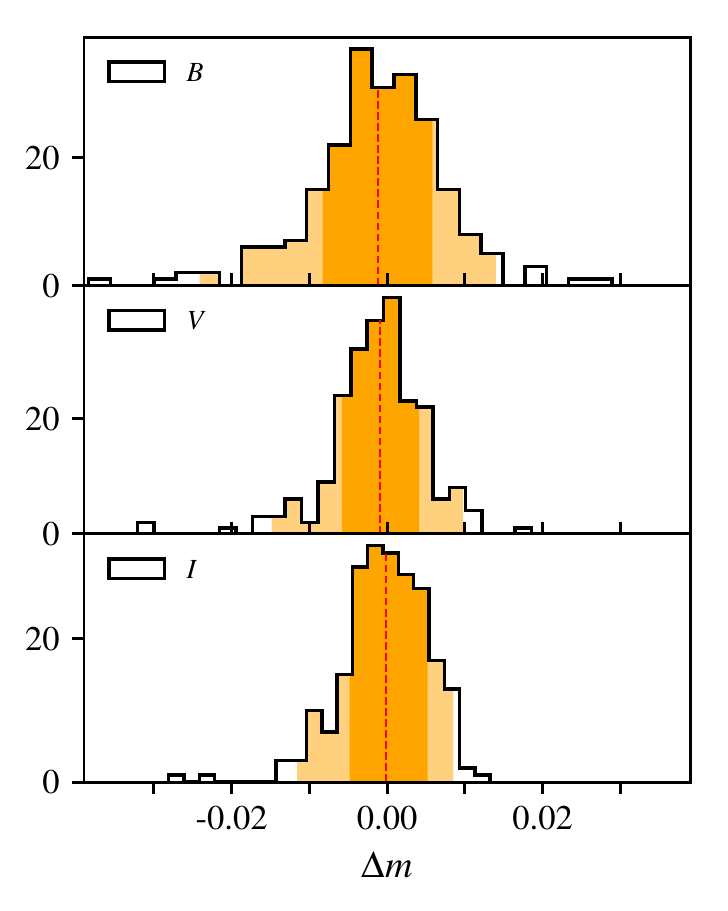}
  \end{center}
  \caption[]{Differences $\Delta m$ between predicted and true magnitudes for 
  the \textit{B}, \textit{V} and \textit{I} bandpasses. The median 
  of each distribution is marked with a dashed red line. We indicate the 
  central 68\% and 95\% intervals in orange and light orange respectively.}
  \label{fig:mag_validation}
\end{figure}

\begin{figure}
  \begin{center}
  \hspace*{-0.45cm}
  \vspace{-0.45cm}
    \includegraphics[]{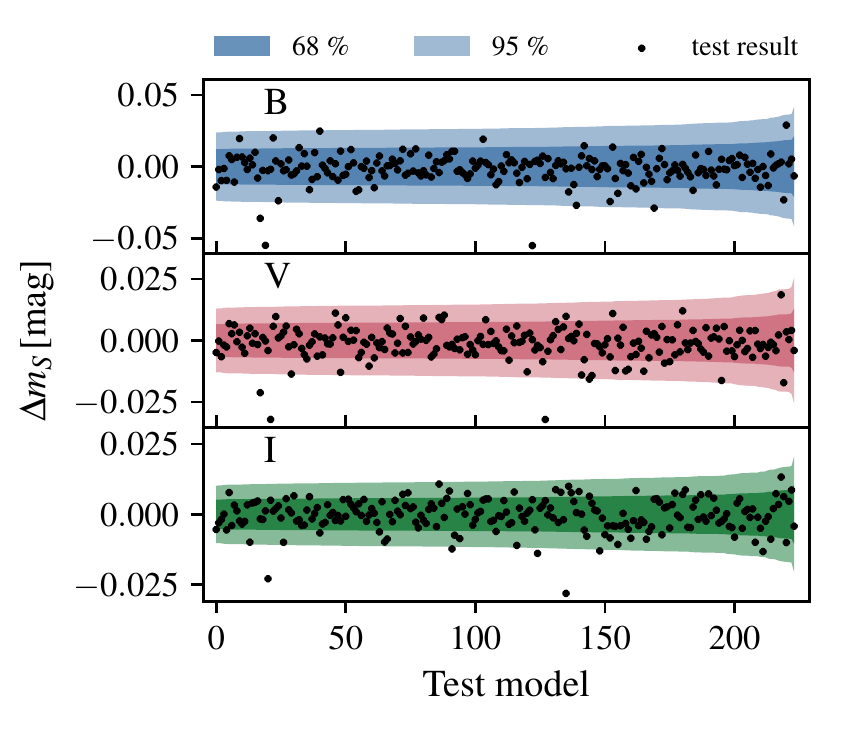}
  \end{center}
  \caption[]{Comparison of the predicted uncertainties to the actual 
  differences, $\Delta m$, between 
  predicted and true magnitudes. For each bandpass (\textit{B}, \textit{V} and 
  \textit{I}), we show the 68\% and 95\% confidence interval for the predicted 
  magnitudes as well as their actual deviations from the magnitudes of the 
  test models (denoted `test result').}
  \label{fig:mag_uncertainties}
\end{figure}

For the purpose of measuring accurate distances, it is crucial that 
we can accurately predict the luminosity for any combination of input 
parameters. We assess the accuracy of our approach by comparing the synthetic 
photometry of the test models to the absolute magnitudes predicted by the 
emulator. As shown in \cref{fig:mag_validation}, the median difference between 
the predicted and true magnitudes is less than 0.0012 mag; this confirms that 
the emulator provides an unbiased estimate of the true model luminosity. 
The accuracy of the predictions decreases from the redder to the bluer 
bandpasses but is nevertheless excellent in all cases; the slight decrease can 
be easily explained by the different amounts of line blanketing in each filter.
In all filters, 68 \% of the models show differences of less than 0.007 mag 
corresponding to errors in the model flux of less than 0.7 \%. For 95 \% of 
the models, the errors are less than 0.02 mag yielding maximum flux errors of 
around 1.8 \%. Thus, in virtually all cases, the accuracy of the emulator is 
much higher than the uncertainties in most real photometric data.
Finally, in \cref{fig:mag_uncertainties}, we demonstrate that, as in the case 
of spectra, the emulator provides sensible estimates for the predictive 
uncertainties.

\section{Learning behavior of the emulator}
In this section, we address questions about the number of models needed for a 
desired accuracy, the adequacy of the adopted methods, and how the emulator 
compares to the standard approach of picking the best-fitting model from the 
grid. 

\begin{figure}
  \begin{center}
  \hspace*{-0.45cm}
  \vspace{-0.45cm}
    \includegraphics[width=0.45\textwidth]{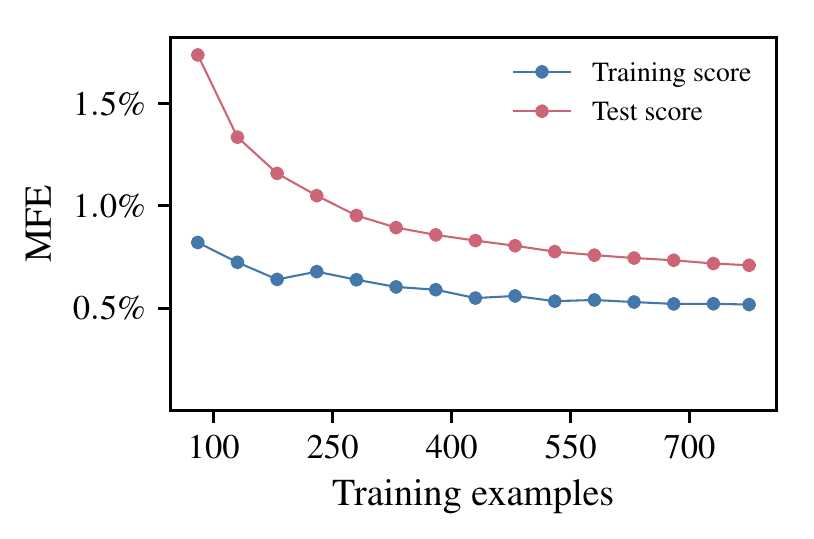}
  \end{center}
  \caption[]{Learning curve for the spectral emulator. The error on the 
  training sample as well as the test sample is shown as a function of the 
  number of models used in the training. The quoted errors are the mean of the 
  individual errors, which, in turn, are the MFE for each spectrum. For each 
  training size, we select five different realizations at random and average 
  the resulting errors.}
  \label{fig:learning_curve}
\end{figure}

We start by creating a learning curve for the spectral emulator 
as shown in \cref{fig:learning_curve}. The learning curve shows the 
accuracy of the emulator as a function of the number of models used for 
training, for both the test and the training sample. 
We keep the number of principal components fixed, thus starting with a minimum 
training size of 80. 
In the investigated regime, the mean error on the training set is almost 
constant at around 0.5 \%. At least part of this error can plausibly
be attributed to the MC noise inherent to the models.\footnote{The MC noise 
manifests itself not only as Poisson noise in the synthetic spectra but also 
in terms of complicated correlated noise that arises from the MC 
uncertainties in the plasma state quantities.}
From the small training errors, we conclude that our model does not suffer 
from high bias, that is to say, the model is flexible enough to provide a 
satisfactory fit to the training data. The mean test error decreases steadily 
from its initial value of 1.7 \% as the number of training instances is 
increased and quickly drops below 1 \%. Finally, for the maximum 
training size, a test score of 0.7 \% is reached. At this point, the 
difference between training and test score is small but non-negligible. 
The gap between the scores will be reduced even further
as more training instances are added since the test score is still 
decreasing (albeit at a slower rate). We conclude that our model 
generalizes well and does not overfit the training data.

\begin{figure}
  \begin{center}
  \hspace*{-0.45cm}
  \vspace{-0.45cm}
    \includegraphics[width=0.45\textwidth]{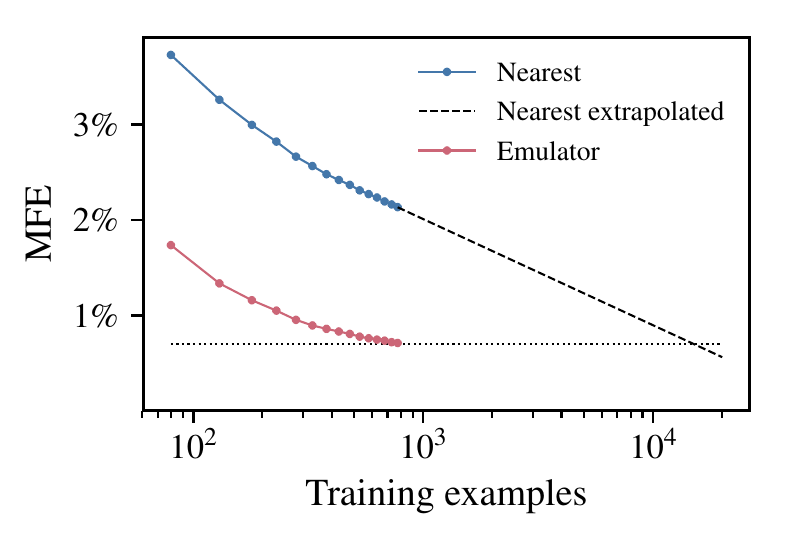}
  \end{center}
  \caption[]{Learning behavior of the emulator as a function of the number of 
  spectra used in the training in comparison to the more naive approach of 
  picking the closest spectrum.}
  \label{fig:nearest}
\end{figure}

Finally, we compare the emulator to the often used approach of simply picking 
the best fitting model from the grid. In \cref{fig:nearest}, we show the test 
scores for both approaches as a function of the number of training instances; 
the plot highlights the massive reduction in the number of models that are 
needed to achieve a given precision. The emulator with the minimum considered 
training size of 80 outperforms 
the method of picking the nearest model even when the full set of {\Ntrain} 
spectra is used. To get a rough estimate of how many models would be needed to 
match the final accuracy of 0.7 \% of the emulator, we linearly extrapolate 
the learning curve in log-linear space; this yields on the order of 15000 
spectra. This is a conservative lower limit for the number of needed models 
since it generously assumes a constant learning rate. 

\section{Modeling observations} \label{sec:observations}
With the spectral emulator, we can fit SN~II spectral time series 
in an automated fashion. For a first demonstration, 
we select SN~1999em and SN~2005cs as our test objects. SN~1999em is 
considered by many to be the prototype of 
a type II supernova, whereas SN~2005cs is a more peculiar, subluminous object. 
Both are among the best-observed type II supernovae, with extensive datasets 
including photometry and spectroscopy at UV, optical and infrared wavelengths 
\citep{Leonard2001,Leonard2002,Hamuy2001,Dessart2006,Pastorello2006,Pastorello2009,Tsvetkov2006,Bufano2009}.
Both SNe have been studied 
with detailed NLTE radiative transfer models using \textsc{Cmfgen} 
\citep{Dessart2006,Dessart2008} and 
\textsc{Phoenix} \citep{Baron2004,Baron2007}. 
We will compare the results of our automated fits to these analyses, which have been conducted carefully
by hand.
In our comparison, we focus on the studies of \citet{Dessart2006} 
and \citet{Dessart2008}, which model more epochs and have published the 
relevant inferred parameters,
namely $T_\mathrm{ph}, v_\mathrm{ph}$ and $n$. In the final step, we infer 
distances to the supernovae from our fits
using the tailored-expanding-photosphere method. 

\subsection{Likelihood for parameter inference}
We use a standard multi-dimensional Gaussian likelihood function 
for parameter inference.
In this case, the log-likelihood of an observed spectrum 
$f_\lambda^\mathrm{obs}$ with $N_\mathrm{pix}$ spectral bins is given by
\begin{equation}
 \ln{p(f_\lambda^\mathrm{obs}|\theta_\mathrm{SN}, E(B-V))} = -\frac{1}{2} \left(
   \mathrm{R}^\mathrm{T} \mathrm{C}^{-1} \mathrm{R} + \ln \det \mathrm{C} + N_\mathrm{pix} \ln 2 \pi
 \right)
\end{equation}
where
\begin{equation}
R = f_\lambda^\mathrm{obs} - f_\lambda (\theta_\mathrm{SN}, E(B-V)) 
\end{equation}
are the residuals with respect to the emulated, reddened spectrum $f_\lambda$ 
and $C$ is the pixel-by-pixel covariance matrix \citep[e.g.,][]{Czekala2014}.
As before, 
$\theta_\mathrm{SN} = (v_\mathrm{ph}, T_\mathrm{ph}, Z, n, t_\mathrm{exp})$ 
are the parameters of our SN model and $E(B-V)$ is the color 
excess.\footnote{We assume a ratio of total to selective absorption of 
$R_V=3.1$ as appropriate for Milky Way type dust.} The residuals have 
pixel-to-pixel correlations mostly due to imperfections in the model 
calculations \citep[see][]{Czekala2014}. For example, a slight error in the 
ionization balance of a given element will lead to features in the synthetic 
spectrum that are either systematically too weak or too strong, producing 
highly correlated residuals in these regions. If these correlations are not 
accounted for in the covariance matrix $C$, the uncertainties of the inferred 
parameters will be severely underestimated \citep{Czekala2014}. We find 
typical uncertainties for the photospheric temperature of the order of a few 
Kelvin if we only include the interpolation uncertainty and the photon count 
noise; thus, without a good statistical model for the correlated residuals, 
the inferred uncertainties are essentially meaningless.
For the purpose of a first demonstration, we resort to a simple maximum 
likelihood approach with homoscedastic white noise, that is to say, a 
diagonal, constant covariance matrix. Our rationale for using homoscedastic 
white noise instead of a combination of the heteroscedastic photon noise and 
the interpolation uncertainties is that the systematic mismatches between 
model and observation are the dominant error component in most regions; 
ignoring it means that we assign highly variable and essentially meaningless 
weights to different parts of the spectrum.

\subsection{Fitting observed spectra} \label{sec:fitting_method}
We want to compare our framework for parameter inference to the 
quantitative 
spectroscopic analyses of \citet{Dessart2006} and \citet{Dessart2008}\textemdash
detailed NLTE studies conducted carefully by hand by experts.
To allow an unbiased comparison of the inferred parameters, we copy key 
assumptions of their study. 
First, we adopt solar metallicities for the non-CNO processed elements. 
Second, we use the same elapsed times since explosion as utilized in the 
calculation of their spectral models; this is particularly important for 
the comparison of photospheric temperatures, which are sensitive to this parameter.
Third, we adopt a color excess of $E(B-V)=0.1$ towards SN~1999em\footnote{This 
value is slightly higher than our favored 
reddening of \\$E(B-V)=0.08$ \citep[see][]{Vogl2019}.} and a color excess of 
$E(B-V)=0.04$ towards SN~2005s in concordance with \citet{Dessart2006} and 
\citet{Dessart2008}. We redden the emulated spectra by this color excess 
according to the \citet*{Cardelli1989} law with $R_V=3.1$.
Finally, we blueshift the observed spectra by the peculiar velocities of their 
host galaxies, which we assume to be 
  \SI{770}{\km \per \second} \citep{Leonard2002,Dessart2006} for SN~1999em and 
  \SI{466}{\km \per \second} \citep{Dessart2008} for SN~2005s.
A log of the spectra used as well as relevant model parameters such as 
the time since explosion utilized in the calculation of the spectral models
can be found in \cref{tab:spec_log}.

\begin{table}
\caption{Log of modeled spectra for (a) SN~1999em, and (b) SN~2005cs.}
\begin{subtable}{0.5\textwidth}
\sisetup{table-format=-1.2}   
\centering
   \begin{tabular}{cccc}
            \hline \hline
      \thead{JD\linebreak\\(+$2\,451\,474.04$)} & Date & Source &${t_\mathrm{exp}}$ \\ 
      \midrule
      17.9 & \space 9 Nov. & \citetalias{Hamuy2001} &9.67 \\ 
      22.9 & 14 Nov. & \citetalias{Hamuy2001} &11.21 \\ 
      27.9 & 19 Nov. & \citetalias{Hamuy2001} &22.00\tablefootmark{a} \\
      \bottomrule
   \end{tabular}
   \caption{SN~1999em}\label{tab:spec_log_99em}
\end{subtable}

\bigskip
\begin{subtable}{0.5\textwidth}
\sisetup{table-format=4.0} 
\centering
    \begin{tabular}{cccc}
      \hline \hline
      \thead{JD\linebreak\\(+$2\,453\,549$)} & Date & Source &${t_\mathrm{exp}}$ \\
      \midrule
      12.25 & 9 Jul. & \citetalias{Dessart2008} & 14.67 \\
      13.5 & 10 Jul. & \citetalias{Faran2014} & 16.12 \\ 
      14.5 & 11 Jul. & \citetalias{Pastorello2006} & 15.33 \\ 
      17.0 & 14 Jul. & \citetalias{Pastorello2006} & 16.11 \\
      19.4 & 16 Jul. & P200 &  19.40 \\
      \bottomrule
   \end{tabular}
   \caption{SN~2005cs}\label{tab:spec_log_05cs}
\end{subtable}
\tablefoot{The reference JDs are the estimated times of explosion of \citet{Dessart2006} and \citet{Pastorello2009}.
 The abbreviations for 
the data sources are \citetalias{Hamuy2001} for \citet{Hamuy2001}, 
\citetalias{Pastorello2006} for \citet{Pastorello2006}, 
\citetalias{Dessart2008} for \citet{Dessart2008}, \citetalias{Faran2014} for 
\citet{Faran2014}, and P200 for
spectra taken at the Palomar 200-inch Hale Telescope with DBSP. All spectra 
have been retrieved from the 
\href{http://www.weizmann.ac.il/astrophysics/wiserep/}{WISeREP} archive
\citep{Yaron2012}. We use the listed time since explosions ${t_\mathrm{exp}}$, 
as taken from \citet{Dessart2006} and \citet{Dessart2008}, for calculating 
synthetic spectra with the emulator.
\tablefoottext{a}{As the single exception, we adopt the maximum 
$t_\mathrm{exp}$ of our spectral emulator for the epoch of the 19th of 
November; this is roughly 19\% smaller than the value of 
$t_\mathrm{exp} =27.0$ used by  \citet[][]{Dessart2006}}.} \label{tab:spec_log}
\end{table}

\subsubsection*{SN~1999em}
\begin{figure*}
\centering
\begin{subfigure}[b]{0.45 \textwidth}
    \includegraphics[]{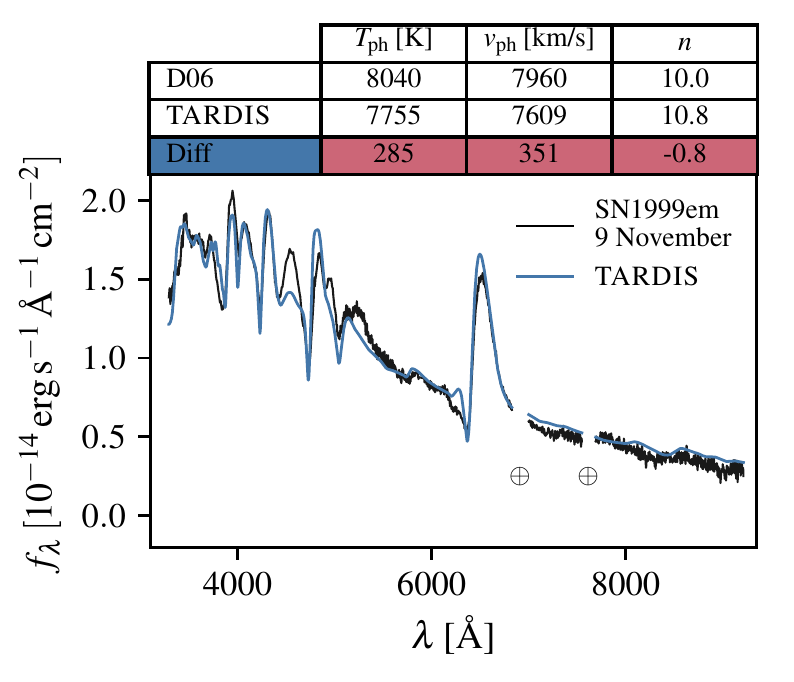}
    \caption{November 9}
   \label{fig:05cs_10}
\end{subfigure}
\begin{subfigure}[b]{0.45 \textwidth}
    \includegraphics[]{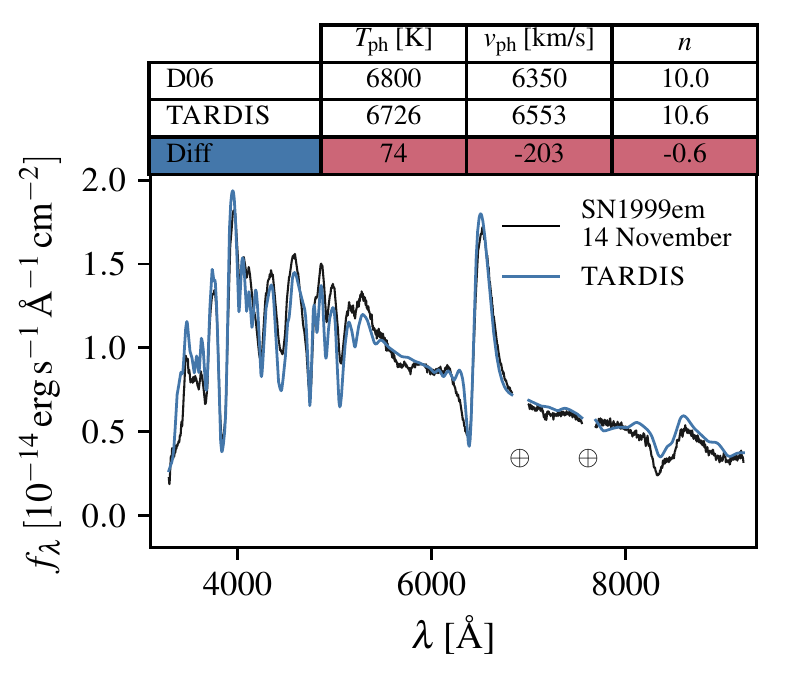}
    \caption{November 14}
   \label{fig:05cs_11}
\end{subfigure} \
\begin{subfigure}[b]{0.45 \textwidth}
    \includegraphics[]{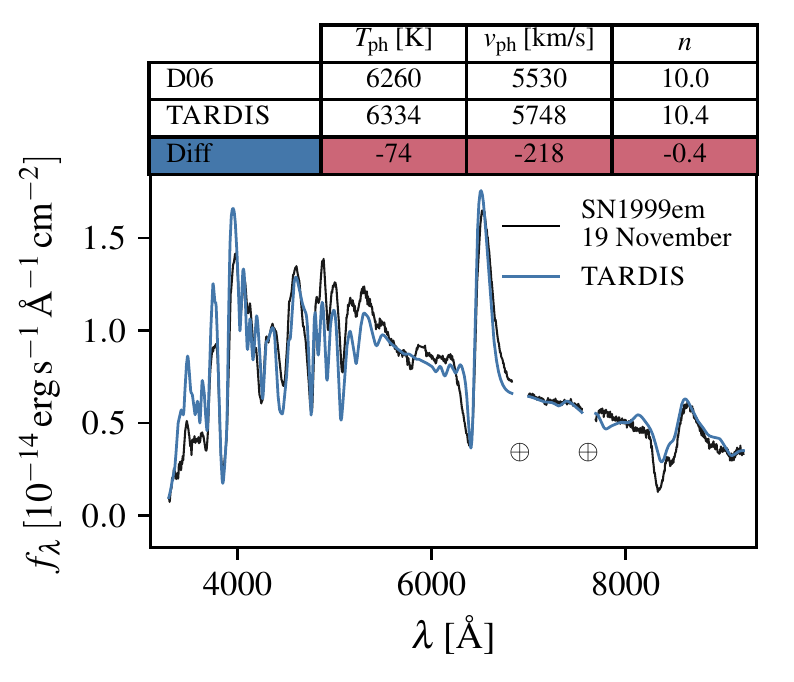}
    \caption{November 19}
   \label{fig:05cs_14}
\end{subfigure}
\\[2ex]
  \caption[]{Spectroscopic analysis of SN~1999em. The three subfigures show 
  comparisons between observed (black) and best-fit emulated spectra (blue); 
  the best-fit has been determined through a maximum likelihood approach as 
  outlined in \cref{sec:fitting_method}. Each spectral comparison is combined 
  with a table of the inferred parameters, the literature values from 
  \citet{Dessart2006} (D06) and the difference between the two. 
 Since the observations have not been corrected for telluric absorption, we 
 exclude the regions of strongest absorption from the fit (marked $\oplus$).}
  \label{fig:fits_99em}
\end{figure*}

\begin{figure}
  \begin{center}
  \hspace*{-0.45cm}
  \vspace{-0.45cm}
    \includegraphics[width=0.45\textwidth]{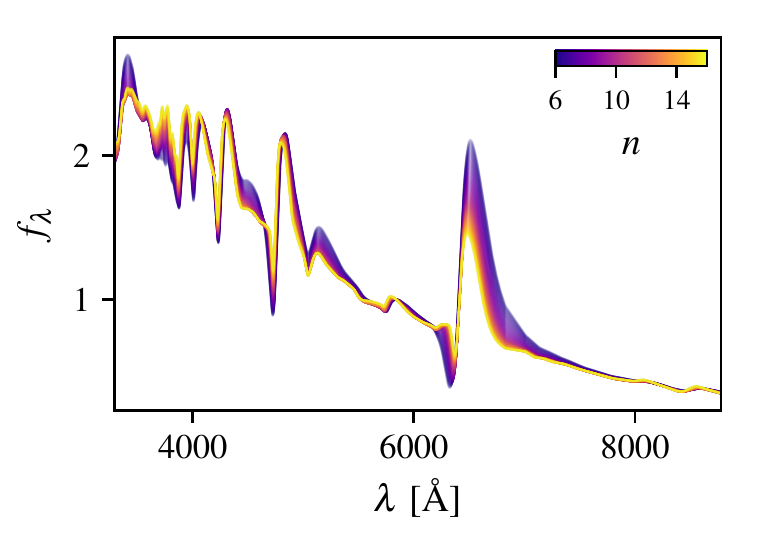}
  \end{center}
  \caption[]{Variation of the spectral shape with the steepness of the density 
  profile $n$. We plot specific flux $f_\lambda$ in arbitrary units 
  for a sequence of emulated spectra where only the power law density 
  index $n$ is modified between spectra; the remaining parameters have been 
  chosen to provide a good fit to SN~1999em for the 9th of November (see 
  \cref{fig:fits_99em} a). We color code the plotted spectra by the power law 
  density index $n$.}
  \label{fig:spec_sequence_n}
\end{figure}

We model three epochs of SN~1999em, covering a time span between 
roughly two to four weeks after explosion. In 
\cref{fig:fits_99em}, we show the maximum likelihood emulated spectra
in comparison to the observations, highlighting the good agreement between the 
two.\footnote{Nevertheless, even better agreement between models and 
observations would be achieved for our favored reddening of 
$E(B-V)=0.08$ \citep[see][]{Vogl2019}.} For each spectral epoch, a table 
with the inferred maximum likelihood parameters as well as the literature 
values from \citet{Dessart2006} is attached to the plot. Despite using vastly 
different methods for calculating synthetic spectra 
and for adjusting them to match the observations, we find good agreement 
in the inferred parameters, with maximum differences of only
$285\,$K in photospheric temperature, $351\,$km/s in photospheric velocity and 
$0.8$ in the steepness of the density profile. We visualize this in 
\cref{fig:param_comp},
which plots our values for $v_\mathrm{ph}$ and $T_\mathrm{ph}$
against those of \citet{Dessart2006}.
Whereas our best fit parameters for $T_\mathrm{ph}$ and $v_\mathrm{ph}$
fall below or above those of \citet{Dessart2006} depending on the epoch,
we find systematically higher values for the power law density index $n$. 
To investigate this, we examine the influence of the steepness of the density 
profile on the emergent spectra in \cref{fig:spec_sequence_n} using the epoch 
of the 9th of November as an example. From this, it becomes clear that in the 
discussed regime of values (density indexes between 10 and 11) the changes in 
the emergent spectra are small. 
We see that only strong lines such as H$\alpha$, which form over a wide range 
of velocities, are affected at all and even those only slightly.

\subsubsection*{SN~2005cs}
\begin{figure*}
\centering
\begin{subfigure}[b]{0.45 \textwidth}
    \includegraphics[]{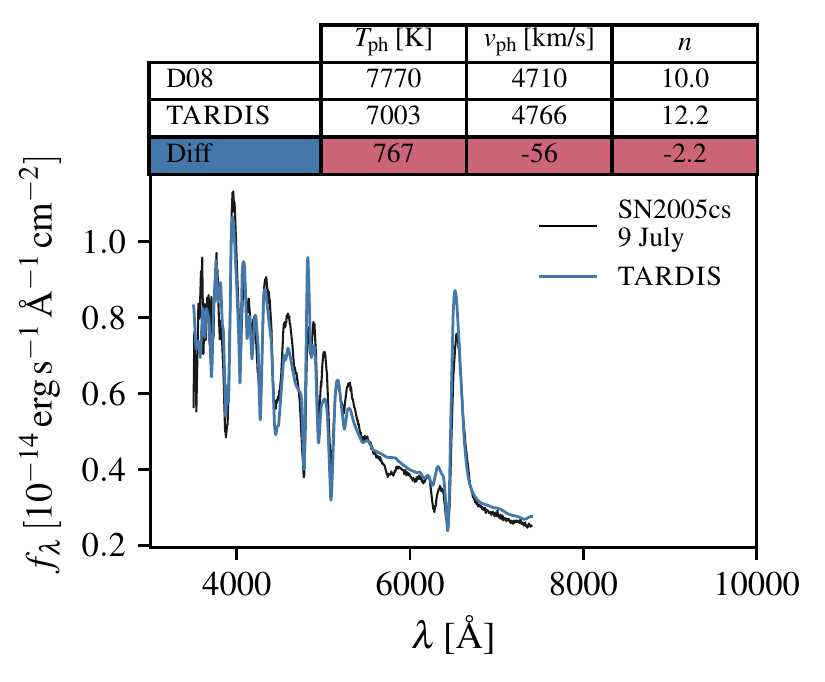}
    \caption{July 9}
   \label{fig:05cs_14}
\end{subfigure}
\begin{subfigure}[b]{0.45 \textwidth}
    \includegraphics[]{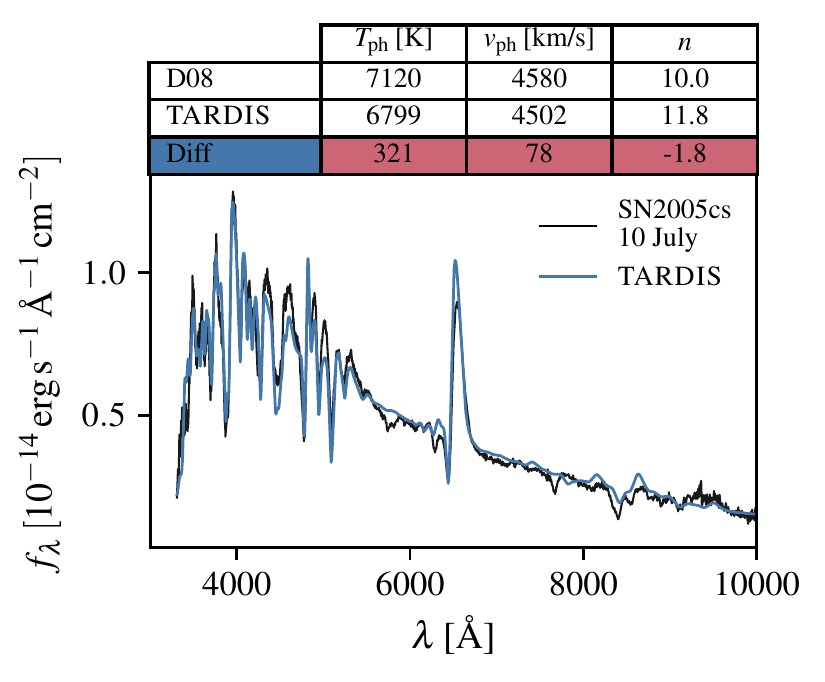}
    \caption{10 July}
   \label{fig:05cs_10}
\end{subfigure}
\begin{subfigure}[b]{0.45 \textwidth}
    \includegraphics[]{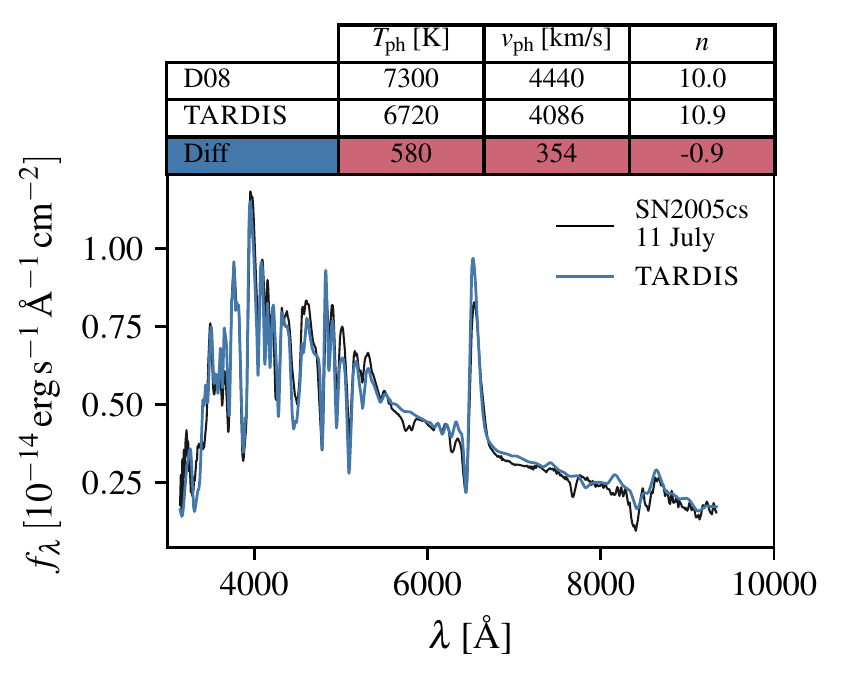}
    \caption{11 July}
   \label{fig:05cs_11}
\end{subfigure} 
\begin{subfigure}[b]{0.45 \textwidth}
    \includegraphics[]{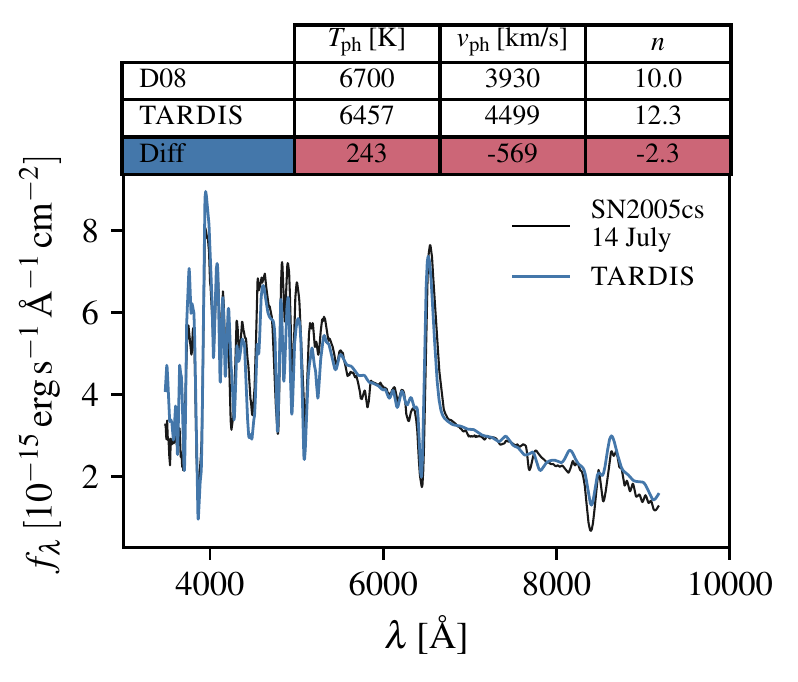}
    \caption{July 14}
   \label{fig:05cs_14}
\end{subfigure}
\begin{subfigure}[b]{0.45 \textwidth}
    \includegraphics[]{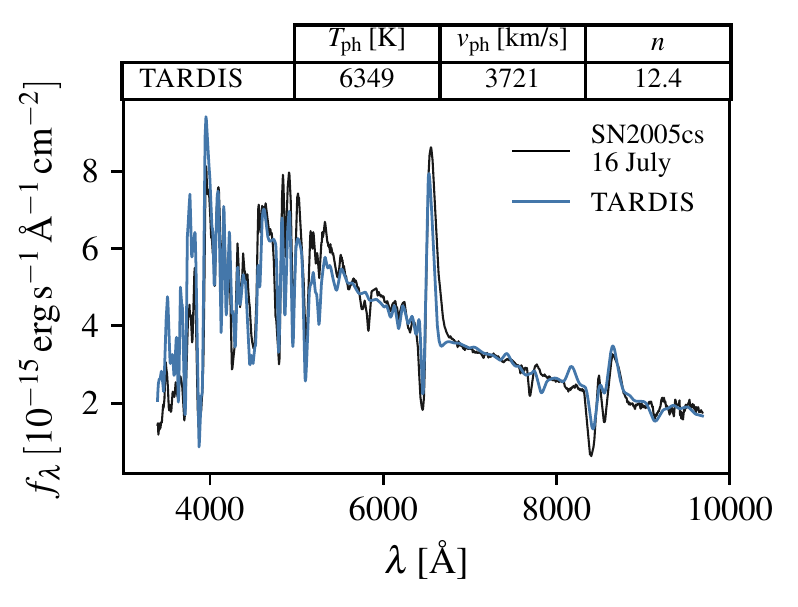}
    \caption{July 16}
   \label{fig:05cs_10}
\end{subfigure}
\\[2ex]
  \caption[]{Spectroscopic analysis of SN~2005cs. See \cref{fig:fits_99em} for 
  a description of the layout. Here, values for the parameter comparison are 
  taken from \citet{Dessart2008} (D08). In (e), we only show the inferred 
  parameters since this epoch has not been modeled in \citet{Dessart2008}.}
  \label{fig:fits_05cs}
\end{figure*}

We analyze five closely-spaced spectral observations of SN~2005cs, between 
roughly two to three weeks after explosion. For the first four epochs, there 
are spectral models from \citet{Dessart2008} at comparable epochs, allowing a 
comparison of the inferred parameters. The last epoch on the 16th of July is 
used only for our measurement of the distance to the supernova in 
\cref{sec:Distances}.
As for SN~1999em, we show the maximum likelihood emulated spectra 
combined with tables of the inferred and literature parameters in 
\cref{fig:fits_05cs}. Again, we find good agreement in the inferred parameters 
with only few exceptions.

For the photospheric velocity, the epoch of the 14th 
of July stands out, which shows a deviation of $569\,$km/s. In particular, the 
increase in velocity compared to the previous epoch is 
puzzling. It can be
understood in the following way:
as discussed in \citet[][\S 3.3.3]{Dessart2006}, 
spectral fits yield differences on the \SI{10}{\percent} level in the photospheric
velocity depending on which set of lines the fit is optimized on.
For the two epochs, our automated fits 
likely attribute varying weights to certain features, thus giving rise to the inconsistent velocity
estimates; for example,
on the 11th the Ca infrared triplet absorption
is not fit well, forming at too low velocities, whereas on the 14th the absorption minimum
is matched much better.

\begin{figure*}
  \begin{center}
  \hspace*{-0.45cm}
    \includegraphics[width=\textwidth]{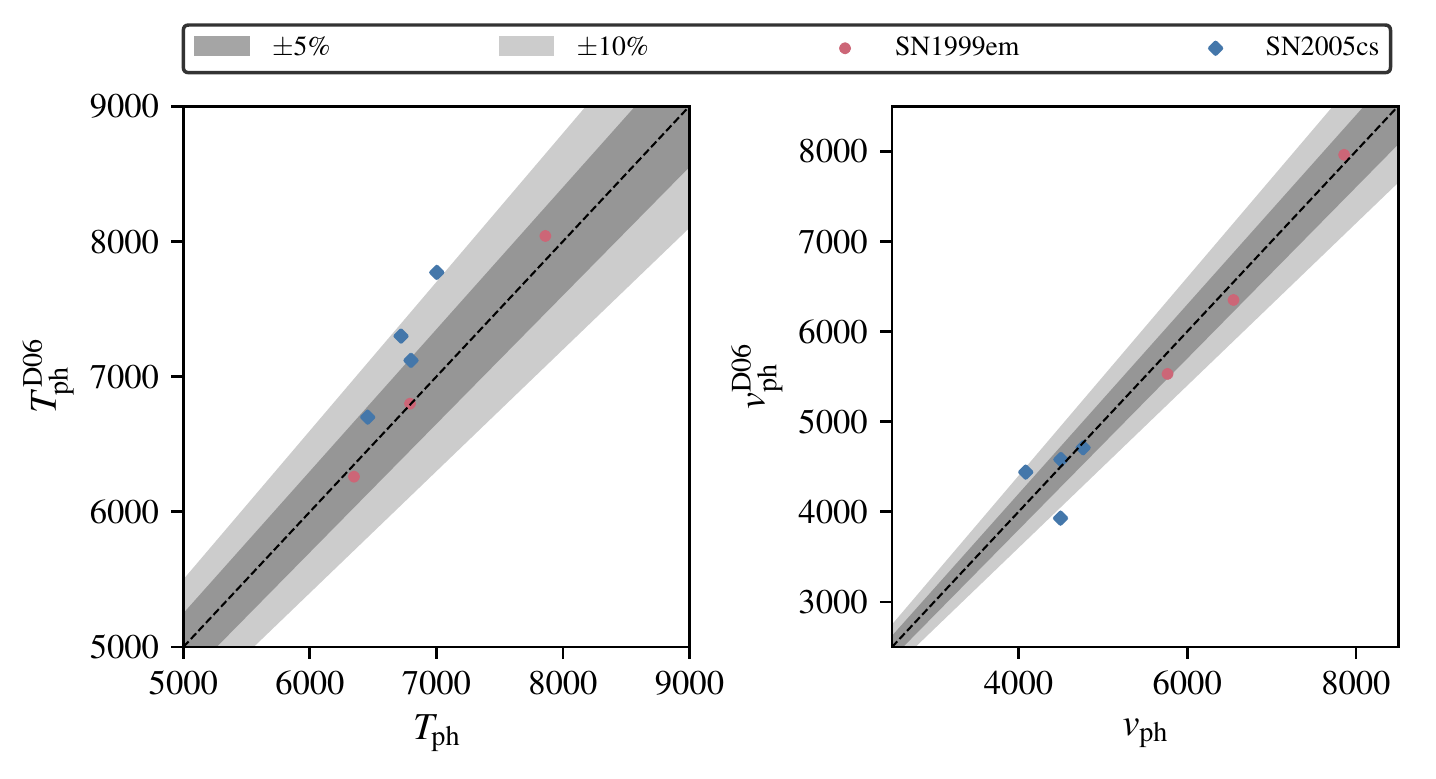}
  \end{center}
  \caption[]{Comparison of the photospheric temperatures $T_\mathrm{ph}$ and 
  velocities $v_\mathrm{ph}$ 
  inferred from our automated spectral fits to those of 
  \citet{Dessart2006} and \citet{Dessart2008}: $T_\mathrm{ph}^\mathrm{D06}$ 
  and $v_\mathrm{ph}^\mathrm{D06}$. The dashed black line indicates perfect 
  agreement between the measurements, whereas the grey shaded 
  regions denote deviations of 5 \% and 10 \% respectively.}
  \label{fig:param_comp}
\end{figure*}

Regarding photospheric temperature, the earliest epoch has the largest 
deviation, which is $767\,$K.
We do not know for certain what causes this significant difference.
Nevertheless, it is striking that the epoch with the largest deviation in 
temperature also has the smallest wavelength coverage.
We show a full comparison of measured temperatures and velocities in 
\cref{fig:param_comp}.

Similar to the case of SN~1999em, our maximum likelihood fits 
favor slightly steeper density profiles than those proposed by 
\citet{Dessart2008}; instead of $n=10$, we find values between 10.9 and 12.4. 
As outlined in the previous section, these variations in the density profile 
only induce very moderate changes in the emergent spectrum and should not be 
overinterpreted.\footnote{We note that \citet{Baron2007} have invoked 
similarly steep density distributions at even later epochs (see their 
radiative transfer model for the 31st of July).} This applies, in particular, 
to the increase of the best-fit value for $n$ in the last two epochs; this is 
likely not due to a physical effect but an artifact of our current method of 
using different density profiles for each epoch and our maximum likelihood 
approach. This will be alleviated by fitting the entire spectral time 
series at the same time.

\subsection{Distance measurements} \label{sec:Distances}
In the past, the need to optimize the fit quality by hand and eye 
combined with the high cost of radiative transfer calculations 
have made distance measurements from SN~II spectral models
\citep[e.g.,][]{Baron2004,Dessart2006} a 
very labor-intensive process. Automated fits based on spectral emulation
revise this picture completely.

We use a variant of the tailored EPM \citep{Dessart2006,Dessart2008} to 
constrain the distance $D$ to the supernovae. As a first step, we 
measure the photospheric angular diameter 
$\Theta = R_\mathrm{ph} / D = v_\mathrm{ph} t_\mathrm{exp} / D $ 
for each epoch.
We compare the apparent magnitudes $m_S$ of our best fit model
\begin{equation}
  m_S = M_{S}(\theta^*) - 5 \log(\Theta) + A_S
\end{equation}
to the observed photometry ${m_S}^\mathrm{obs}$ for different 
values of $\Theta$.
Here, $A_{S}$ is the broadband dust 
extinction for the bandpass \textit{S}=\{\textit{B},\textit{V},\textit{I}\} 
and $M_{S}(\theta^*)$ 
is the predicted absolute magnitude for the best fit parameters $\theta^*$. 
We adopt the photospheric angular diameter $\Theta^*$ that 
minimizes the squared difference between observed and model magnitudes:
\begin{equation}
\Theta^* = \underset {\Theta}{\operatorname {arg\,min}} \sum_S \left(m_S - {m_S}^\mathrm{obs}\right)^2
\end{equation}
Our approach to tailored EPM is technically identical to \citet{Dessart2006} 
and \citet{Dessart2008} but avoids the detour of parametrizing the 
model magnitudes by a blackbody color temperature and a dilution factor.

Finally, we determine the time of explosion and the distance through 
a Bayesian linear fit to the
time evolution of the ratio of the photospheric angular diameter $\Theta$ and 
the photospheric velocity $v_\mathrm{ph}$.
To be more specific, we obtain the time of explosion from the intercept with 
the time axis and the distance from the inverse of the slope. In our analysis, 
we assume that the uncertainties are Gaussian and that they have standard 
deviations of 10 \% of the measured values as in \citet{Dessart2006} and 
\citet{Dessart2008}.\footnote{In the case of a full Bayesian analysis, the 
assumption of Gaussian uncertainties can be dropped 
and the posterior distribution of $\Theta / v_\mathrm{ph}$ can be used 
instead.}
We use a half-Cauchy prior for the slope of the regression curves,
corresponding to a uniform distribution for the angle between the straight 
lines and the time axis. We adopt informative priors for the time of 
explosion, which we will discuss in the context of the 
individual supernovae below.

Our sources of photometry are \citet{Leonard2002} for SN~1999em \citep[as 
listed in Table 1 of][]{Dessart2006} and
\citet{Pastorello2009} for SN~2005cs.
If there is no coincident photometric observation for a given 
spectral epoch, we linearly interpolate the magnitudes from the nearest 
epochs. We list all magnitudes used 
in our tailored EPM analysis in \cref{tab:photometry}.

\begin{table}
\caption{Interpolated \textsc{BVI} photometry for the epochs of spectral 
observations.}
\begin{subtable}{0.5\textwidth}
\sisetup{table-format=-1.2}   
\centering
   \begin{tabular}{ccccc}
      \toprule \toprule
      \thead{JD\linebreak\\(+$2\,451\,474.04$)} & Date & \textsc{B} & \textsc{V} & \textsc{I} \\ 
      \midrule
        17.9 & \space 9 Nov. & 14.02 & 13.84 & 13.48 \\ 
        22.9 & 14 Nov. & 14.25 & 13.81 & 13.44 \\ 
        27.9 & 19 Nov. & 14.47 & 13.86 & 13.40 \\
      \bottomrule
   \end{tabular}
   \caption{SN~1999em}\label{tab:photometry_99em}
\end{subtable}

\bigskip
\begin{subtable}{0.5\textwidth}
\sisetup{table-format=4.0} 
\centering
    \begin{tabular}{ccccc}
      \toprule \toprule
      \thead{JD\linebreak\\(+$2\,453\,549$)} & Date & \textsc{B} & \textsc{V} & \textsc{I} \\ 
      \midrule 
        12.25 &9 Jul. &   14.75 &  14.59 &  14.30 \\
        13.5 &10 Jul. &  14.83 &  14.60 &  14.26 \\
        14.5 &11 Jul. &  14.92 &  14.58 &  14.25 \\ 
        17.0 &14 Jul. &  15.09 &  14.67 &  14.25 \\ 
        19.4 &16 Jul. &  15.26 &  14.70 &  14.28 \\
      \bottomrule
   \end{tabular}
   \caption{SN~2005cs}\label{tab:photometry_05cs}
\end{subtable}
\tablefoot{The reference JDs are the same as in \cref{tab:spec_log}.}
\label{tab:photometry}
\end{table}

\subsubsection*{SN~1999em}
For a first demonstration of the emulator, we adopt an informative Gaussian 
prior for the time of explosion
based on the tailored EPM analysis of \citet{Dessart2006}, 
which finds $t_0$ = JD 2\,451\,474.04$\pm1.0$. While the time of explosion for 
SN~1999em is not well constrained through the photometry, 
many objects have limits that are as tight or tighter than the adopted prior 
for $t_0$; this applies, for example, to SN~2005cs as we will discuss in the 
next section.

\begin{figure*}
\centering
    \includegraphics[width=\textwidth]{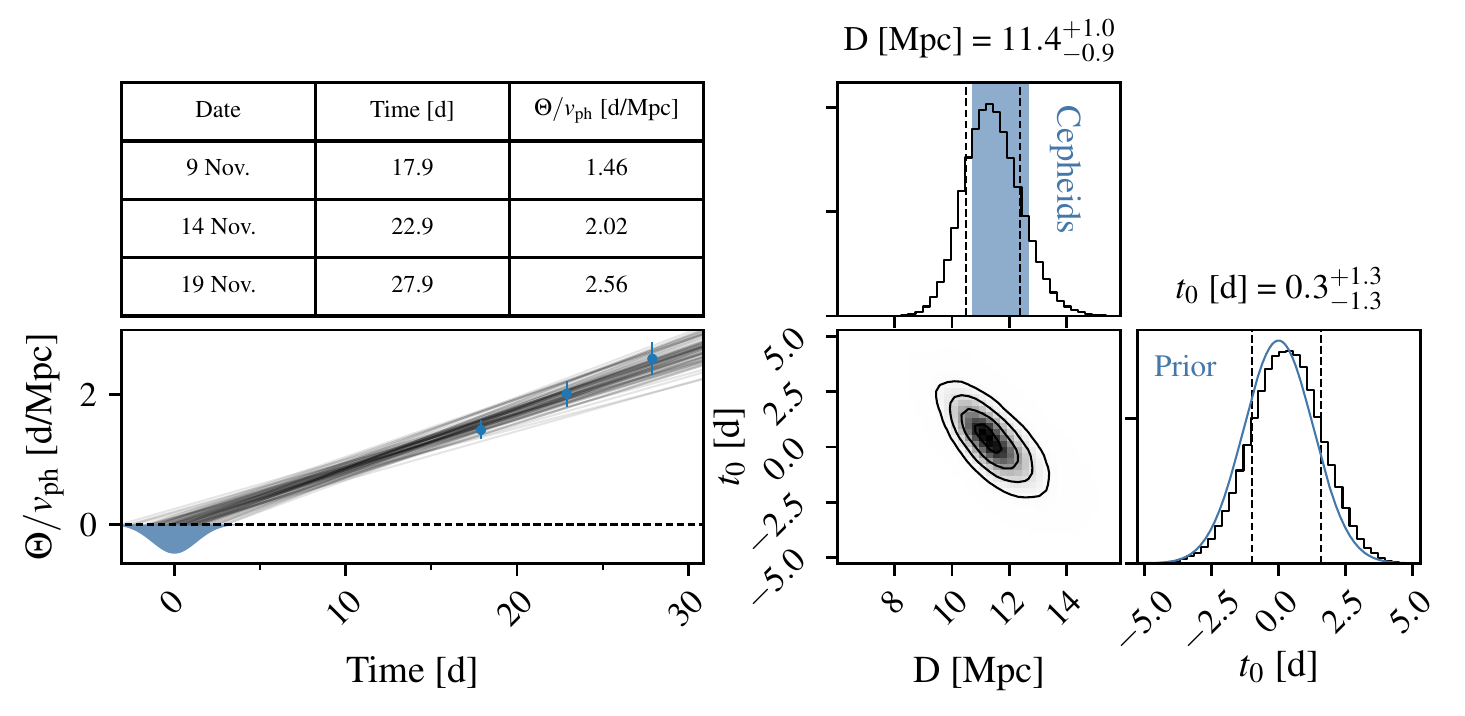}
  \caption[]{Tailored EPM for SN~1999em. The lower left panel shows the 
  evolution of the ratio of the photospheric angular diameter $\Theta$ and the 
  photospheric velocity $v_\mathrm{ph}$ with time (blue circles).
  Here, we measure the time with respect to JD $2\,451\,474.04$. 
  We tabulate the plotted values in the upper left panel.
  Finally, we perform a Bayesian linear fit to this data. Our prior for the 
  time of explosion is indicated by the blue shaded region. From the posterior 
  distribution, we show 100 randomly sampled regression curves for 
  illustrative purposes. The right half of the figure features a corner plot 
  of the inferred parameters. Our distance measurement is in excellent 
  agreement with the Cepheid distance of \citet{Leonard2003}, which is 
  indicated by the blue shaded region.} 
  \label{fig:tepm_99em}
  \vspace*{\floatsep}
  \includegraphics[width=\textwidth]{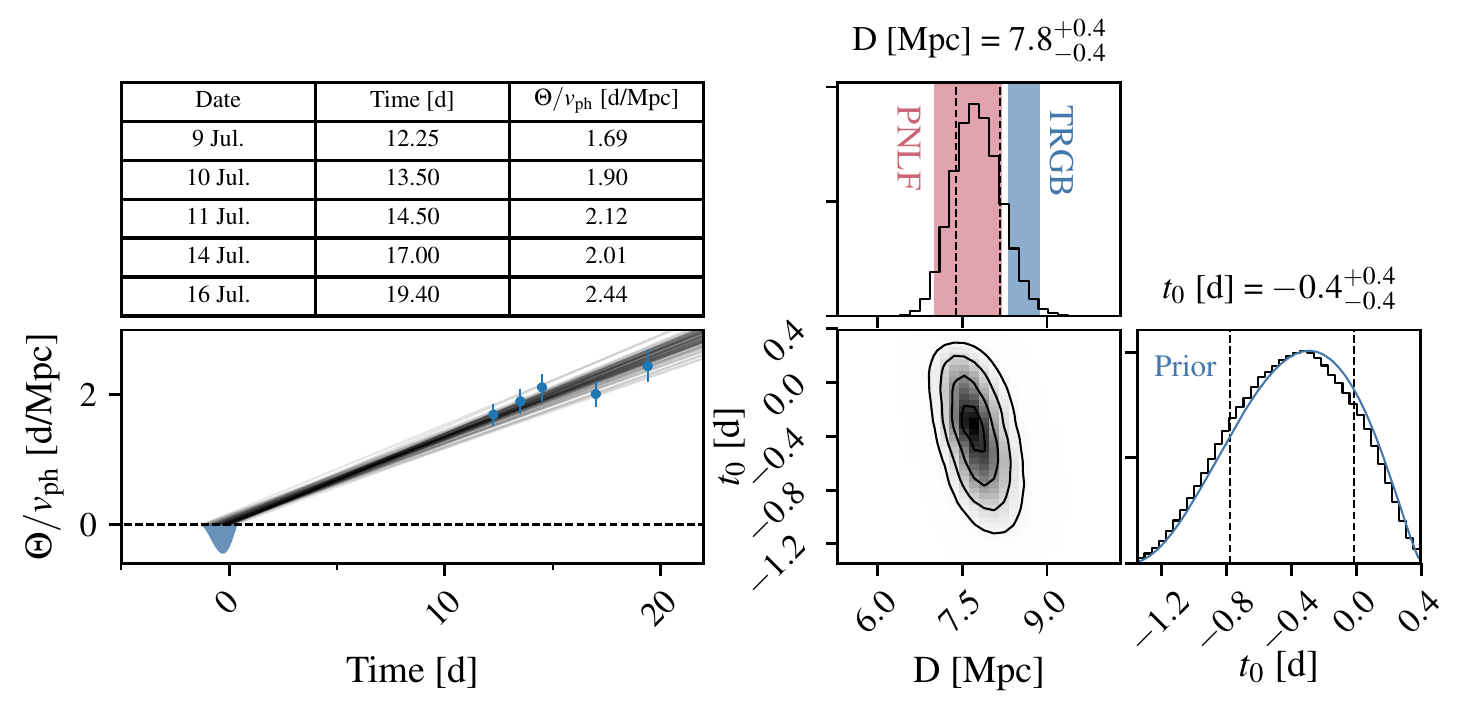}
  \caption[]{Tailored EPM for SN~2005cs. For a description of the figure 
  layout see \cref{fig:tepm_99em}. The time zero point is at JD 
  $2\,453\,549.0$. Within the statistical uncertainties, the inferred distance 
  agrees with the measurement of \citet{Ciardullo2002} using the planetary 
  nebula luminosity function (PNLF), as well as that of \citet{McQuinn2016} 
  based on the tip of the red giant branch (TRGB).}
    \label{fig:tepm_05cs}
\end{figure*}

With the prior for $t_0$ defined, we apply the tailored EPM as outlined above. 
We summarize the inputs as well as the results of our analysis in 
\cref{fig:tepm_99em}. In the figure, we combine a table of the ratios of 
photospheric angular diameter and velocity $\Theta / v_\mathrm{ph}$, a 
visualization of the Bayesian linear regression, and a corner plot of the 
inferred distance and time of explosion. We find a distance of 
${11.4}^{+1.0}_{-0.9}\,$Mpc, which is in excellent agreement with the Cepheid 
distance to the host galaxy
of ${11.7}\pm 1\,$Mpc \citep{Leonard2003}. It is important to keep in 
mind that the quoted uncertainties are solely statistical and depend both on 
the adopted prior for $t_0$ and the assumed uncertainties for 
$\Theta / v_\mathrm{ph}$. 
Finally, we point out that the regression is only weakly informative on the 
time of explosion, that is to say, the posterior distribution for $t_0$ is 
only modified slightly compared to the prior.

\subsubsection*{SN~2005cs}
As opposed to SN~1999em, the time of explosion for SN~2005cs
is constrained tightly by photometric observations.
Based on the non-detection at JD = $2\,453\,548.43$ and
the detection at JD = $2\,453\,549.41$,
\citet{Pastorello2009} identify JD = $2\,453\,549.0 \pm 0.5$ as the
time of shock breakout. In our prior for $t_0$, we make small changes to this 
result to incorporate two basic arguments. First, the prior probability that 
the first detection is coincident with the explosion should be zero. Secondly, 
the probability that the explosion occurred before the last non-detection 
should be non-negligible due to the limited depth of the image.
Based on these considerations, we construct the Beta prior shown in 
\cref{fig:tepm_05cs}. Our prior peaks shortly after the last non-detection and 
has a width that is compatible with the quoted uncertainties of 
\citet{Pastorello2009}. We derive the distance to the supernova as illustrated 
in \cref{fig:tepm_05cs} and obtain a value of ${7.8}^{+0.4}_{-0.4}\,$Mpc.

In contrast to SN~1999em, the distance to SN~2005cs is not constrained 
through Cepheids. 
NED\footnote{The NASA/IPAC Extragalactic Database (NED)
is operated by the Jet Propulsion Laboratory, California 
Institute of Technology,
under contract with the National Aeronautics and Space Administration.} lists 
50 individual distance measurements spanning a range of values between 
$2.45\,$Mpc and $12.2\,$Mpc with a median distance of $7.935\,$Mpc. Based on 
spectral modeling of SN~2005cs,
\citet{Dessart2008} find a distance of $8.9 \pm 0.5\,$Mpc using tailored EPM 
in the \textit{BVI} bandpasses and \citet{Baron2007}  
${7.9}^{+0.7}_{-0.6}\,$Mpc with the SEAM method. Independent state-of-the-art 
measurements come from
\citet{Ciardullo2002}, who derive a distance of $7.6 \pm 0.6\,$Mpc using the 
planetary nebula luminosity function (PNLF), and  \citet{McQuinn2016}, who 
infer a value of $8.58 \pm 0.10\,$Mpc\footnote{The reported uncertainty only 
accounts for statistical errors. The authors speculate that the systematic 
uncertainty could be of order 0.05 mag in distance modulus.} from the 
tip-of-the-red-giant-branch method.

Overall, the agreement between our measurement and the results above is 
satisfactory given the uncertainties of the individual methods. The 15\% 
deviation to the tailored EPM distance of \citet{Dessart2008}, however, 
warrants investigation.
We find that roughly half of the discrepancy arises from differences in the 
time of explosion. From the evolution of the ratio of photospheric angular 
diameter and velocity, \citet{Dessart2008} obtain an explosion epoch that is 
earlier than ours by about a day. They explain the difference between their 
estimate and those based on non-detections 
\citep[specifically,][in their paper]{Pastorello2006} with a short time delay 
between the beginning of the expansion and the optical brightening.
After adjusting the time of explosion,
the remaining deviation is around 7 \% and thus within the expected range.

\section{Conclusions and outlook} \label{sec:conclusions}
In this paper, we demonstrate the use of spectral emulation to predict the 
SN~II spectra and magnitudes generated by \textsc{tardis} 
\citep{Kerzendorf2014,Vogl2019}.

The key ingredient of our approach is the creation of a low-dimensional space 
for the interpolation of synthetic spectra through the combination of 
appropriate data preprocessing and dimensionality reduction by PCA. 
In this space, we train Gaussian processes to predict preprocessed, 
dimensionality-reduced spectra for new sets of input parameters. In the final 
step, we reverse the preprocessing procedure to obtain a spectrum that can be 
compared to observations.
This method emulates the output of our radiative transfer code to high 
precision; we demonstrate this by comparing emulated and simulated spectra for 
a large number of test models. On average, the emulator prediction deviates 
from the simulation by around {\EmuAcc} (as measured by the MFE)\textemdash 
this is much smaller than both observational and model uncertainties. Not only 
are the interpolation uncertainties small but we can also estimate them 
sensibly through our use of Gaussian processes; this will allow us, in the 
future, to propagate these errors into the uncertainties of the inferred 
parameters.

We complement the spectral emulator with an emulator for absolute magnitudes; 
we have discarded the luminosity information
in the spectral emulator
to standardize the spectra and to obtain better predictive performance.
The training data are
Johnson-Cousins \textit{B},\textit{V}, \textit{I} magnitudes that have been 
synthesized from the unprocessed training spectra. We remove variations in the 
luminosity that result from differences in the physical sizes of the supernova 
models by transforming the magnitudes to the position of the photosphere; 
then, we interpolate the transformed magnitudes using Gaussian processes.
This allows us to predict absolute magnitudes with an average precision of 
better than 0.01 mag, which is significantly smaller than typical 
observational uncertainties.

The emulator is not only accurate, but also orders of magnitude faster 
($\approx10\,$ms) than our simulator 
\textsc{Tardis} ($\approx100000\,$s)\textemdash this makes it possible to 
fit spectra automatically. We demonstrate this by performing maximum 
likelihood parameter estimation for spectral time series of SN~1999em and 
SN~2005cs. The inferred parameters of the supernovae show good agreement with 
those of \citet{Dessart2006} and 
\citet{Dessart2008}, who studied these objects 
using the \textsc{Cmfgen} code. Similarly, the distances that we infer 
from our fits are consistent with the best available measurements from the 
literature. 

As a next step, we will develop a more detailed likelihood to infer accurate 
uncertainties for a complete parameter estimate. The emulator and an advanced 
likelihood will then allow the use of type IIP supernova for accurate 
cosmological distance determinations.

\begin{acknowledgements}
CV thanks Andreas Floers, Marc Williamson, and Markus 
Kromer for stimulating discussions during
various stages of this project. The authors 
gratefully acknowledge Stefan
Taubenberger for sharing his enormous expertise in 
the field of supernovae and
for being a key member of the supernova cosmology 
project that motivated this work.
The authors thank the anonymous reviewer for valuable comments.
This work has been supported by the Transregional 
Collaborative Research Center
TRR33 `The Dark Universe' of the Deutsche 
Forschungsgemeinschaft and by the 
Cluster of Excellence `Origin and Structure of the 
Universe' at Munich Technical
University. SAS acknowledges support from STFC 
through grant, ST/P000312/1. This research made use of \textsc{Tardis}, a 
community-developed software
package for spectral synthesis in supernovae
\citep{Kerzendorf2014, kerzendorf_wolfgang_2019_2590539}.
The development of \textsc{Tardis} received support from the
Google Summer of Code initiative
and from ESA's Summer of Code in Space program. \textsc{Tardis} makes
extensive use of Astropy and PyNE.
Data analysis and visualization was performed using 
\textsc{Matplotlib} \citep{Hunter2007},
\textsc{Numpy} \citep{numpy} and \textsc{SciPy} 
\citep{Scipy}.
\end{acknowledgements}

\bibliographystyle{aa} 
\bibliography{Mendeley,tardis,references}

\clearpage

\begin{appendix}
\section{Min-max normalization} \label{sec:MinMax}
Min-max normalization scales each feature (here,  
the fluxes in a bin) individually 
such that it spans the range [Min, Max] for the training data.
The min-max normalized flux $f_{\lambda,ij}^\mathrm{norm}$ for spectrum $i$ in bin $j$ is given by
\begin{equation} \label{eq:MinMax}
    f_{\lambda,ij}^{\mathrm{norm}} = \frac{f_{\lambda,ij} - \underset {i}{\operatorname {min}} f_{\lambda,ij}}{
    \underset {i}{\operatorname {max}} f_{\lambda,ij} - \underset {i}{\operatorname {min}} f_{\lambda,ij}} \left(
    \mathrm{Max} - \mathrm{Min}
    \right) + \mathrm{Min}.
\end{equation}
This linear transformation can be easily reversed in the prediction step of the emulator.
\end{appendix}

\end{document}